\documentclass[twoside]{article}
\usepackage{booktabs}
\usepackage[linesnumbered, ruled,vlined]{algorithm2e}
\SetKwInput{KwInput}{Input}
\SetKwInput{KwOutput}{Output}

\usepackage[accepted]{aistats2024}
\usepackage{amsmath,amssymb,amsfonts}
\usepackage{float}
\usepackage{graphicx}
\usepackage{amsthm}
\usepackage{algorithmic}
\usepackage{graphicx}
\usepackage{textcomp}
\usepackage{caption}
\usepackage{xcolor}
\usepackage{subcaption}
\usepackage{booktabs}
\usepackage{mathrsfs}
\usepackage{url}
\usepackage{bigfoot}

\newtheorem{lemma}{Lemma}
\newtheorem{thm}{Theorem}
\newtheorem{Assumption}{Assumption}

\usepackage[round]{natbib}

\usepackage{multirow}
\usepackage{xr}
\begin{document}
%

%

\twocolumn[


\aistatstitle{MIM-Reasoner: Learning with Theoretical Guarantees for Multiplex Influence Maximization}

\aistatsauthor{Nguyen Do$^{1}$ \And Tanmoy Chowdhury$^{2}$ \And Chen Ling$^{2}$ \And Liang Zhao$^{2}$ \And My T. Thai$^{3,\dagger}$}

\aistatsaddress{
$^1$Posts and Telecommunications Institute of Technology, Ha Noi, Viet Nam \\
$^2$Emory University, Atlanta, USA, \quad $^3$University of Florida, Gainesville, USA\\

}]


\newcommand{\nguyendhk}[1]{\textcolor{blue}{\# #1 (NguyenDHK) \# }}

\newcommand{\mt}[1]{\textcolor{red}{\# #1 (MT) \# }}

\begin{abstract}
Multiplex influence maximization (MIM) asks us to identify a set of seed users such as to maximize the expected number of influenced users in a multiplex network. MIM has been one of central research topics, especially in nowadays social networking landscape where users participate in multiple online social networks (OSNs) and their influences can propagate among several OSNs simultaneously. Although there exist a couple combinatorial algorithms to MIM, learning-based solutions have been desired due to its generalization ability to heterogeneous networks and their diversified propagation characteristics. In this paper, we introduce MIM-Reasoner, coupling reinforcement learning with probabilistic graphical model, which effectively captures the complex propagation process within and between layers of a given multiplex network, thereby tackling the most challenging problem in MIM. We establish a theoretical guarantee for MIM-Reasoner as well as conduct extensive analyses on both synthetic and real-world datasets to validate our MIM-Reasoner's performance.

\end{abstract}

\section{INTRODUCTION}
Many users on Online Social Networks (OSNs), such as Facebook and Twitter, are increasingly linking their accounts across multiple platforms. The multiple linked OSNs with overlapping users is defined as a Multiplex Network. The structure of multiplex networks allows more users to post information across various OSNs simultaneously, offering significant value for marketing campaigns \citep{lim2015cross}. Although this interconnectivity of the multiplex network enables seamless information flow between platforms via overlapping nodes, the underlying information propagation models on each OSN can differ. This means that the way information spreads and influences users on one OSN may not be the same as on another OSN. Given these distinct characteristics, designing a customized influence maximization strategy to wield considerable influence over various platforms becomes imperative.



As a combinatorial optimization problem, Influence Maximization (IM) aims at selecting a small subset of users to maximally spread information throughout the network. In the past decades, tremendous combinatorial optimization algorithms have been proposed on a single network \citep{Kempe,proxybased1,jiang2011simulated,tang2015influence,nguyen2016stop,li2018influence,tang2018online,tiptop,guo2020influence}. They have achieved faster running time with approximation and exact solutions under certain propagation models. In parallel to traditional combinatorial optimization-based methods, machine learning based approaches \citep{li2022piano, li2023survey, chen2022touplegdd, ling2023deep} have also been proposed in recent years and achieved success in handling massive complicated networks and generalizing well to similar problems. Both types of methods cannot be trivially adapted to the multiplex network scenario since they only consider a single model of influence propagation, whereas each network in a multiplex network can have a different propagation model.




To date, a few combinatorial optimization algorithms \citep{zhan2015influence,Zhang_2016,kuhnle2018multiplex,singh2019mim2,katukuri2022cim} have been proposed to tackle Multiplex IM (MIM). However, despite the numerous benefits offered by machine learning-based approaches, the progress of applying such methods to MIM is still in its infancy owing to the following two fundamental challenges. \textit{(1) Scalability.} A multiplex network is of very large size with heterogeneous propagation models, of which each layer has its own characteristics. 
Current state-of-the-art (SOTA) IM learning-based methods face computational limitations, especially in the case of simulating and estimating the propagation spread within and between layers. 
\textit{(2) Generalizability For Lightweight Model.} In multiplex networks, capturing diverse propagation characteristics often requires large GNN-based models. 
However, using such large GNN models significantly increases the time complexity, resulting in longer training times and slower inference processes. Balancing the model size and computational efficiency is crucial when working with multiplex networks.

\textbf{Our Contributions.} To overcome both scalability and generalization for lightweight model issues altogether, we propose a novel framework MIM-Reasoner that decomposes multiplex networks into individual layers and leverages deep reinforcement learning to find near-optimal seed nodes for the multiplex network as a whole.  
Specifically, we first employ the Knapsack approach to assign appropriate budgets to each layer, while aiming to maximize the overall spread of the whole multiplex network, thereby obtaining a $(1-\epsilon)$-decomposition. With such a decomposition strategy, 
MIM-Reasoner learns a lightweight policy to find feasible solutions for each layer and minimize the overall computational workload and complexity of the whole network. Through the learning process, MIM-Reasoner models the interdependent relationship between the current layer and other layers, which avoids reactivating nodes 
and maximizes the overall spreading process. In addition, we establish guarantees for the solutions discovered by the optimal policy of MIM-Reasoner in the worst-case, best-case, and general scenarios. We empirically demonstrate the strength of MIM-Reasoner in both synthetic and real-world datasets from the perspective of influence spread and running time.

\section{PROBLEM STATEMENT AND BACKGROUND}
    A multiplex network consisting of $k$ layers is represented by 
    $\mathscr{G}=\{\left(G_1, \sigma_1\right), \ldots,\left(G_k, \sigma_k\right)\}$, where each element consists of a directed graph $G_i=\left(V_i, E_i\right)$, and an influence model $\sigma_i$ (i.e Independent Cascade (IC) or Linear Threshold (LT) \citep{Kempe}) that describes the propagation of influence within $G_i$. If a node belongs to multiple layers, an interlayer edge is added between corresponding nodes 
        to signify the node's overlapping presence. The entire set of nodes in the multiplex is denoted by $V$ 
    where $V=\bigcup_{i=1}^k V_i$. 
    Without loss of generality, we can assume that $V_i=V$ for
    all $i$. 
    If a vertex $v \in G_i$ does not exist in some $G_{\eta}$, we can simply add it to $G_{\eta}$ as an isolated vertex. In this work, as we allow each layer of a multiplex network to have a different model of influence propagation, it becomes necessary to establish a mathematical definition for the propagation model on 
    $\mathscr{G}$.
    
    \textbf{Definition 1} (\textit{Influence Propagation Model)} \citep{kuhnle2018multiplex}.  A model of influence propagation, denoted as $\sigma_i$, on a graph $G_I=(V, E_i)$ is defined by a function $P$ that assigns probabilities to the final activated sets $T \subset V$ given a seed set $S \subset V$. The probability $P(T \mid S) \in [0,1]$ satisfies the property $\sum_{T: T \subset V} P(T \mid S) = 1$, ensuring that we have a probability distribution. The expected number of activated nodes, denoted as $\sigma_i(S)$, given a seed set $S$, is calculated as follows:
    
    \begin{equation}
    \sigma_i(S) = \sum_{T: T \subset V} P(T \mid S) \cdot |T|
    \end{equation}
    
    The above definition covers most of the propagation models in the literature, such as IC, LT, and SIR \citep{Kempe}.
    
    The influence propagation model $\sigma$ on  $\mathscr{G}$ is defined as follows: if an overlapping node $v$ is activated in one 
    layer graph $G_i$, then its adjacent interlayer copies in other layers also become activated in a deterministic manner, called {\em overlapping activation}. The propagation of influence occurs independently within each graph $G_i$ according to its respective propagation model $\sigma_i$.
    Note that we count the duplicated nodes as a single instance rather than adding up all of them. 
    We are now ready to define our MIM problem as follows:

    

    
    \textbf{Definition 2} ({\bf Multiplex Influence Maximization (MIM)}). Given a multiplex graph $\mathscr{G} = \{(G_1 = (V, E_1), \sigma_1), \dots, (G_k = (V, E_k), \sigma_k)\}$ and a budget $l \in \mathbb{N}$, the MIM problem asks us to find a seed set $S \subset V$ of size at most $l$ so as to maximize the expected number of activated nodes in the multiplex network denoted as $\sigma(S)$. An instance of this problem is denoted as $(\mathscr{G}, k, l, \sigma)$ and finding the optimal set of seed node $\Tilde{S}$ that maximizes spreading among those graphs follows the following function:
    
    \begin{equation}
    \tilde{S}=\arg \max _{|S| \leq l} \sigma \left(S\right)
    \label{eq: problem-definition}
    \end{equation}
    
    For each layer $G_i \in \mathscr{G}$, many greedy based 
    algorithms \citep{leskovec2007cost,CELF++,tang2014influence,tang2015influence} have obtained a performance guarantee bound of  $(1-1/e)$, if $\sigma_i$ is submodular and monotone increasing 
    \citep{Kempe}. If 
    all $\sigma_i$ of all $G_i$ satisfy the Generalized Deterministic Submodular (GDS) property, then $\sigma$ is submodular \citep{kuhnle2018multiplex}. 
    
\begin{figure}[t]
    \centering
    \includegraphics[width=0.7\linewidth]{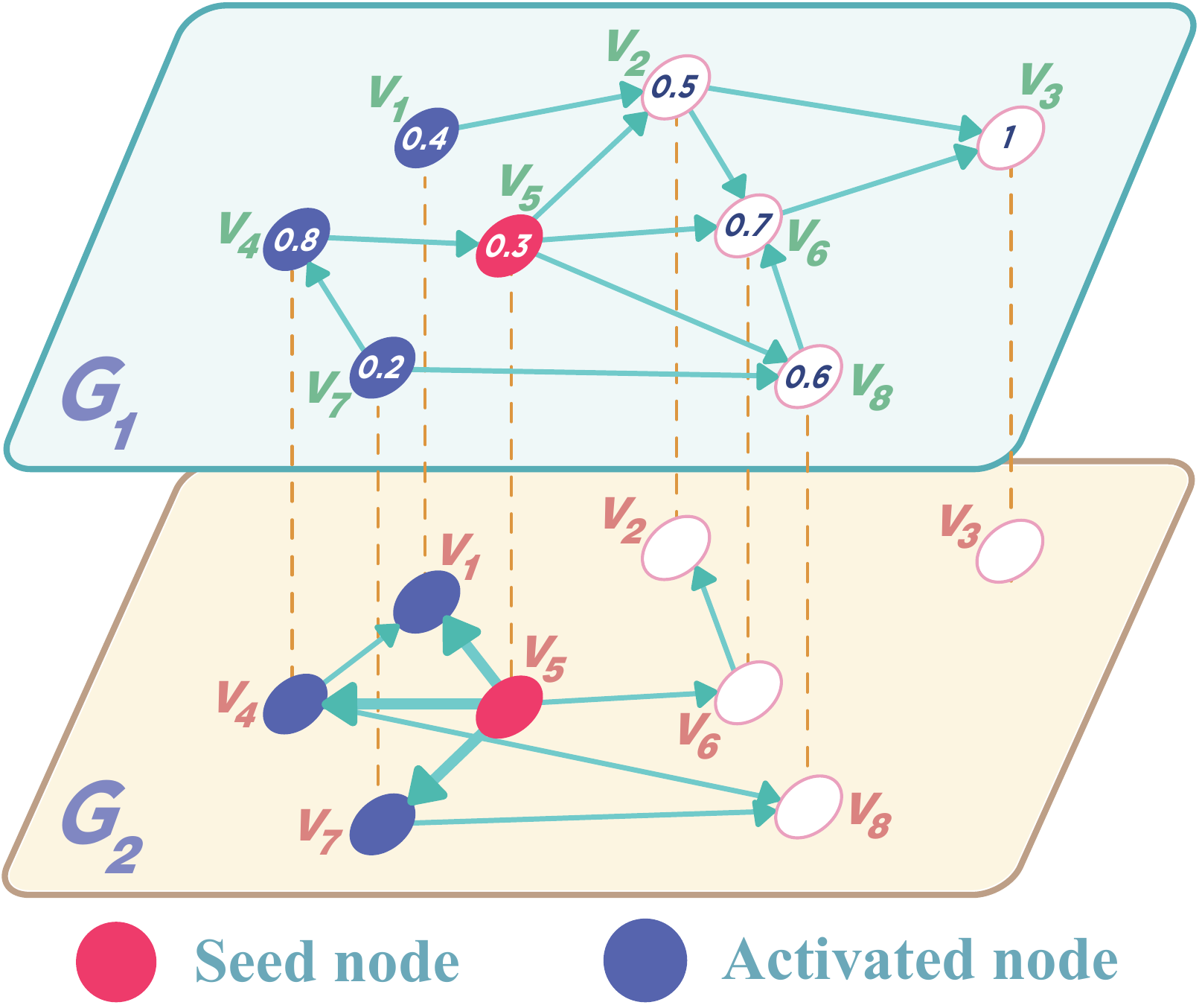}
       \caption{An example of the propagation process in a multiplex network consisting of 2 layers. Layers $G_1$ and $G_2$ operate LT and IC model, respectively. Each node in $G_1$ has a threshold $\zeta \in [0,1]$. The green bold arrow in $G_2$ indicates high probability of activation.}
       \label{fig:problem1_pointing}
    \end{figure}

    To visualize the propagation process in multiplex networks, let's consider an instance of the MIM problem denoted as $(\mathscr{G}, 2, 1, \sigma)$. Here, $\mathscr{G}={(G_1=(V, E_1),\sigma_1), (G_2=(V, E_2),\sigma_2)}$ represents a two-layer multiplex network as shown in Figure \ref{fig:problem1_pointing} 
    where $\sigma_1$ of $G_1$ is an LT model and $\sigma_2$ of $G_2$ follows an IC model. In this instance, node $v_5$ is selected as the seed node. Assuming that in layer $G_2$, $v_5$ activate nodes $[v_1,v_4,v_7]$.  Those nodes in layer $G_1$ are also activated due to the overlapping activation property. Moreover, $v_8$ has a low chance to be activated by other nodes in layer $G_2$. However, in layer $G_1$, node $v_8$ has a threshold $\zeta=0.6$, meaning it requires at least two activated neighboring nodes to be activated itself. In this case, $[v_5,v_7]$ are activated, leading to the activation of $v_8$ in both layers. Similarly, the activation process continues to propagate to other nodes, such as $[v_2,v_6,v_3]$ in layer $G_1$, hence, $[v_2,v_6,v_3]$ in layer $G_2$ are also activated in the deterministic manner. As a result, by properly selecting seed nodes, we only need one budget to activate all nodes in $\mathscr{G}$.
    
\section{RELATED WORK}
\begin{figure*}[t]
\centering
\includegraphics[width=1\linewidth]{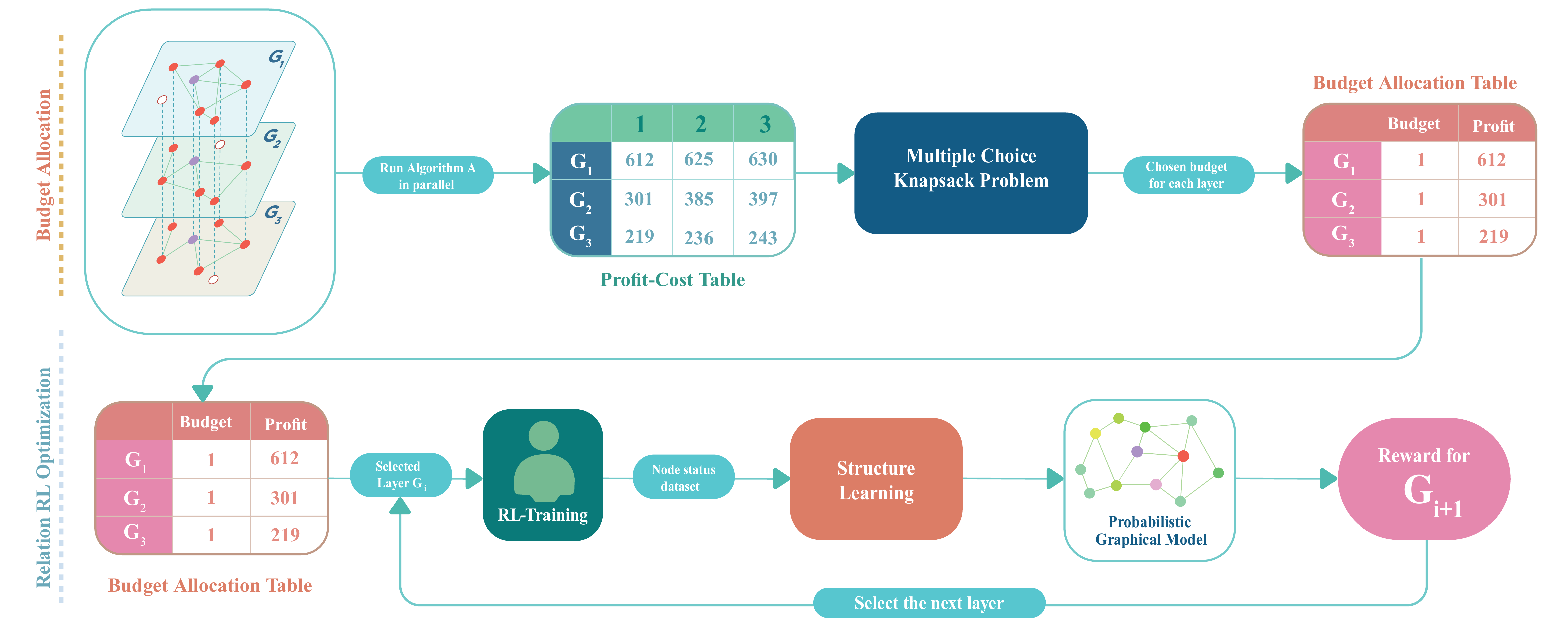}
    \caption{MIM-Reasoner consists of two phases: Budget Allocation and Relation RL Optimization. In Phase 1, algorithm $\mathcal{A}$ calculates the profit and cost for each layer in parallel. The 'Multiple Choice Knapsack Problem' is then solved to determine the allocated budget for each layer, presented in the 'Budget Allocation Table'. In Phase 2, an RL-Agent is trained to sequentially find solutions for each layer using the allocated budget. Simultaneously, the status dataset is processed through 'Structure Learning' to create a 'Probabilistic Graphical Model' which reveals layer relationships and helps the RL-Agent to avoid reactivating nodes already activated by other layers.}
    \label{fig:flow-MIM-Reasoner}
\end{figure*}
\textbf{Combinatorial optimization for IM.} IM, which was firstly introduced by \cite{Kempe}, is a well-studied problem in network analysis. Traditional approaches include simulation-based \citep{leskovec2007cost,CELF++, UBLF}, proxy-based \citep{proxybased1,proxybased2,proxybased3}, and approximation-based methods \citep{li2018influence, jiang2011simulated, tang2015influence,nguyen2016stop,tang2018online,guo2020influence}.
Most of these works obtained a $(1-1/e)$-approximation ratio due to the submodularity of propagation models. Recently, Tiptop \citep{tiptop} has been introduced as an almost exact solution to IM. For more detailed reviews of traditional methods, we refer readers to a recent survey by \cite{banerjee2020survey}. In the context of MIM, the notable {\em approximation} algorithms with theoretical guarantees using combinatorial methods are \citep{Zhang_2016, least-cost-influence, kuhnle2018multiplex}. In \citep{Zhang_2016, least-cost-influence}, 
the authors addressed the Least Cost 
Influence problem
by mapping layers into a single layer using lossless
and lossy coupling schemes. These works assume that
all layers have the same propagation model. Later, \citep{kuhnle2018multiplex} considered the heterogeneous propagation models
and presented Knapsack Seeding of Networks (KSN),
which serves as a starting point for our learning model.

\textbf{ML-based for IM.} The rise of learning-based approaches that utilize deep learning techniques to enhance generalization capabilities has been applied to IM. Reinforcement learning (RL) integrated with IM has shown promise \citep{lin2015learning, ali2018boosting}, with recent state-of-the-art solutions focusing on learning latent embeddings of nodes or networks for seed node selection \citep{manchanda2020gcomb,chen2022touplegdd,li2022piano}. Graph neural networks (GNNs) have also been explored in IM to encode social influence and guide node selection \citep{ling2022source,ling2022deepgar}. Recent methods \citep{ling2023deep, chowdhury4663083deep} have proposed DeepIM as a generative approach to solving IM and achieved state-of-the-art performance. However, these methods only focused on the IM problem, and could not readily be extended to our MIM problem due to the scalability issues and the ability to capture the inter- and intra-propagation relations in a multiplex network. 

\section{MIM-REASONER}
In this section, we introduce our MIM-Reasoner, a deep reinforcement learning (RL) approach coupled with a Probabilistic Graphical Model (PGM) designed to maximize influence in a multiplex network under a budget constraint (Figure \ref{fig:flow-MIM-Reasoner}). In a nutshell, to tackle the scalability issue, MIM-Reasoner, decomposes $\mathscr{G}$ into individual layers, enabling it to find solutions for each individual layer in parallel in the first phase. It utilizes a
knapsack-based approach \citep{mckp} in order to effectively allocate the budget for each layer.
Furthermore, to capture the inter-activation via overlapping users, in the second phase, we train a PGM for each layer. Based on PGMs, MIM-Reasoner then learns a policy $\pi$  that operates sequentially from one layer to another. The PGMs themselves suggest the rewards that can aid policy $\pi$ further tuning the solutions found in each layer to obtain our final solution within a bound of $\frac{(1-\epsilon)(1 - 1/e)}{(o + 1)k}$.

\subsection{Phase 1. Budget Allocation}
This phase aims to find an initial candidate seed set solution serving as a starting point for our deep RL agent in Phase 2. In particular, the key point is how to allocate the budget for each layer so that it can maximize the overall expected number of activated nodes while solving each layer individually. Firstly, this phase involves running any algorithm $\mathcal{A}$ that solves the IM problem  (e.g., a Greedy-based algorithm) with budget $l$ for each layer $G_i$ to measure $\sigma(S_{ij})$ where $j \in [0,...,l]$. 
Note that $G_i$ can be in billion-scale. Thus, Let's denote $\mathcal{E}_i = [\hat{G}_{i,1},...,\hat{G}_{i,h}]$ as a set of subgraphs sampled from $i^{th}$ layer of $\mathscr{G}$; i.e. $G_i$. Here, $h$ represents the total number of sampled subgraphs of $i^{th}$ layer. We adopt a Graph Attention Network (GAT), denoted as $\mathcal{I}(.)$ on a set of subgraphs $\mathcal{E} = [\mathcal{E}_1, \mathcal{E}_2, \cdots, \mathcal{E}_k]$ by following a probabilistic greedy concept (\cite{manchanda2020gcomb}) (details in Appendix \textbf{A.1}).
This trained GAT helps identify a good candidate set $V_i^g \subset V$ before running 
$\mathcal{A}$, thereby further reducing the search space and improving efficiency. 

Given a feasible seed node set $S_{ij}$ found by $\mathcal{A}$ for each $G_i$, MIM-Reasoner constructs a Profit-Cost table $\mathcal{H}$ that records the profit $p(S_{ij})=\sigma(S_{ij})$ associated with the overall spreading achieved in $\mathscr{G}$, while the cost $c(S_{ij})=|S_{ij}|$ indicates the corresponding budget $j$ spent to achieve that spreading. We next solve the Multiple Choice Knapsack Problem (MCKP) to allocate budgets effectively for each layer. In the MCKP, we aim to maximize the total profit $\sum_{i=1}^k p\left(S_{ij}\right)$ while satisfying the constraint $\sum_{i=1}^k c\left(S_{ij}\right) \leq l$. For $\epsilon > 0$, MCKP solver \citep{mckp} has a ($1-\epsilon$) approximation for arbitrary $\epsilon > 0$. The output of MCKP is the budget allocation table $\mathcal{U}^{k \times 2}$, where each row corresponds to a layer $G_{i}$, and the two columns represent the spent budget and profit, respectively.

\subsection{Phase 2. Relation RL Optimization}

In this core phase, 
MIM-Reasoner's goal is to train the policy $\pi$ to optimize our obtained solution, given the budget allocation table $\mathcal{U}$, achieved in Phase 1. Phase 2 has a key component which is a PGM (\cite{PGMs}) denoted as $\varphi$ which is specifically a Bayesian Network trained to capture the complex propagation process within and between layers to provide a novel reward signal to the RL agent. This reward signal assists the agent in finding a nearly optimal seed node set for the current layer while avoiding the reactivation of nodes already activated by previously selected layers. 

To begin, let's define $\mathcal{G}^{prev}=[\varphi_1,...,\varphi_{k-1}]$ as a set that contains the trained PGMs for each layer in a multiplex network. Initially, $\mathcal{G}^{prev} = \emptyset$ since no layer has been selected to learn yet. At each step, a layer $G_i$ is selected to train the policy $\pi$ based on the order provided by $\mathcal{U}$ as follows:
\begin{equation}
G_i = \underset {i \in {1,...,k}}{\operatorname{argmin}} \ \mathcal{U}_{i,2}
\label{eqn:layer_selection}
\end{equation}
Given an allocated budget $j \leq l$, MIM-Reasoner starts with the layer having the lowest profit, say $G_{\hat{i}}$. After selecting the starting layer $G_{\hat{i}}$ for the second phase, it removes row $\hat{i}$ from $\mathcal{U}$ to ensure that $G_{\hat{i}}$ will not be selected in the future. Given layer $G_{\hat{i}}$, it trains the policy $\pi$ to improve the seed node set $S_{\hat{i}j}$, where $j=\mathcal{U}_{\hat{i},1}$, representing the allocated budget. During the policy training process, a status dataset $\mathcal{D}=[0,1]^{m \times |V|}$, where $m$ represents the number of Monte Carlo simulation steps, is also collected. This dataset records the activation status of nodes in the currently selected layers and also indirectly records the activation of nodes in other layers due to the overlapping activation.

Next, MIM-Reasoner trains a PGM $\varphi_{\hat{i}}$ for the selected layer $G_{\hat{i}}$ using a step called Structure Learning, where the learned structure is a directed graph modeling the causal relationship between trained nodes. It takes $\mathcal{D}$ as input and returns a learned structure as output for $\varphi_{\hat{i}}$ 
(details in Appendix \textbf{A.3}). Note that, during the Structure Learning of $\varphi_{\hat{i}}$, only a subset of representative nodes are trained to reduce the complexity, other nodes that have high correlation with each other are grouped together but they can be inferred based on their Pearson Correlation \citep{cohen2009pearson} with trained representative nodes. Details for this Variable Grouping are presented in Appendix \textbf{A.2}. 

After training $\varphi_{\hat{i}}$, it is added to $\mathcal{G}^{prev}$. Since $\varphi_{\hat{i}}$ is trained on $\mathcal{D}$, which records the activation of nodes in other layers, it can predict the relationships between nodes in other layers. Suppose in the next 
step, the layer with the lowest influence $G_{\eta}$ ($\eta \neq \hat{i}$) is chosen using Equation \ref{eqn:layer_selection}. This time, MIM-Reasoner utilizes $\mathcal{G}^{prev}$ to generate rewards that assist the policy $\pi$ in selecting a seed set which ensures that nodes activated by the previously selected layers (i.e., $G_{\hat{i}}$) are not reactivated. These reward functions are discussed next in the (RL) framework.

\subsection{Reinforcement Learning Framework}

This section presents our RL framework, which includes state representation, action space, and reward function. Moreover, we also explain how PGM $\varphi$ helps the RL agent to effectively maximize the overall spread within the allocated budget for each layer of $\mathscr{G}$. 

\textbf{State Representation}: 
We leverage $\mathcal{I}(.)$ trained in Phase 1, to predict a set of candidate nodes $V^g \subset V$ for an unseen graph $\mathscr{G}$ and even further reduce the complexity of the state space and action space. Note that $V_i^g$ represents the set of candidates that potentially contain optimal or near-optimal seed nodes in $G_i$. The state space, $C_{i,t} = (V_i^g, X_i
, S_{i,t}, t) $, describes the state of $G_i$ at time step $t$
while $X_i$ is the feature vector of layer $G_i$ extracted by Structure2vec \citep{dai2016discriminative}, $S_{i,t}$ is the partially computed solution at time step $t$, and $V_i^g$ represents the set of candidate nodes predicted from the pretrained GAT, $\mathcal{I}(.)$. 

\textbf{Action Space}: In the context of our problem, at each time step $t$, the policy $\pi$ will select an action that corresponds to a seed node $v \in V_i^g \backslash S_{i,t}$. 

\textbf{Reward Function}. Recall in the Variable Grouping step, we used our proposed Node Grouping algorithm to identify highly correlated node groups denoted as $\mathcal{P} = [P_1, P_2, \ldots, P_q]$ which satisfies $V=\bigcup_{i=1}^q P_i$ (Details in Appendix \textbf{A.2}). 
Our grouping algorithm automatically divides the nodes into $q$ clusters based on correlation without human involvement. Each group $P_i$ contains nodes that exhibit high correlation with each other. We define the set of representative nodes by taking the node closest to the centroid from each group as $\mathcal{Y} = [v_1, \ldots, v_q]$, where each representative node $v_i \in \mathcal{Y}$ corresponds to a group $P_i \in \mathcal{P}$. During the training process, if the agent identifies a new seed node $v_n$ and adds it to a partially computed solution $S_{i,t}$, the updated set $S_{i, (t+1)}=S_{i,t} \cup v_n$ can activate a subset of nodes $T \subseteq V$. It is important to note that each node $u \in T$ has its own representative node. 

For convenience in determining which group a node $u \in T$ belongs to and which node is the representative node for $u$, we define a surjective function $f: T \longrightarrow \mathcal{Y}$, where for all nodes $\hat{v} \in \mathcal{Y}$, there would be at least an element $u \in T$ such that $f(u) = \hat{v}$. Suppose the policy $\pi$ is currently trained on layer $G_{i}$, we define an activation score function that allows every node $u \in T$ to be based on the representative node set $\mathcal{Y}$ to infer its score for being activated by any other selected layer $G_{\hat{i}}$ using PGMs. This function can be defined as:


\begin{equation}
\mathcal{W}(u)=Q_{u, f(u)} \cdot \underset{\varphi_{\hat{i}} \in \mathcal{G}^{p r \kappa v}}{\operatorname{argmax}} \varphi_{\hat{i}}(f(u),\mathcal{K})
\end{equation}

Here, $Q_{u, f(u)} \in [-1,1]$ represents the Pearson correlation between node $u$ and its corresponding representative node $f(u)$. Meanwhile, $\mathcal{K} = \mathcal{Y} \cap T$ refers to the set of activated representative nodes. On the other hand, in layer $G_{i}$, the output of $\varphi_{\hat{i}}(f(u),\mathcal{K})$ represents the conditional probability of representative node $f(u)$, (i.e. $ P_{\hat{i}}(f(u) \mid \mathcal{K})$) being activated by other layer $G_{\hat{i}}$, given that the current set of activated representative nodes is $\mathcal{K}$. Putting $\mathcal{K}$ as evidence to PGM indicates how likely the representative node $f(u)$ is to be activated by other layers when certain other representative nodes are already active. Based on activation score $\mathcal{W}$, we now present our customized evaluation function $\mathcal{M}(.)$ which measures the total spread in layer $G_i$, given solution $S_i$. It can be defined as:

\begin{equation}
\mathcal{M}(S_i) = \sum_{i=1}^{|T|} [1 -\mathcal{W}(u)]
\label{eqn:spreadingevaluation}
\end{equation}

In Equation \ref{eqn:spreadingevaluation}, if a node $u$ has a high activation score, it means that $u$ is more likely to be activated by previously selected layers. Thus, instead of counting the contribution of $u$ to the total spread as 1, we can modify it to be $1 - \mathcal{W}(u)$. This means that if the reward function is based on $\mathcal{M}$ to calculate marginal gain, the policy $\pi$ has less reward when activating node $u$, hence, it avoids finding a seed set that re-activates nodes that have been already activated by other layers. Finally, we provide our reward function to train policy $\pi$ computed by considering the marginal gain of adding node $v_t$ to the partially computed solution $S_t$, as follows:

\begin{equation}
r_t = \mathcal{M}(S_{i, (t+1)}) - \mathcal{M}(S_{i, t})
\label{eqn:rewardfunction}
\end{equation}

Here, $\mathcal{M}(S_{i, (t+1)})$ is the total spread given a partially computed solution at time step $t+1$, and $\mathcal{M}(S_{i,t})$ represents the spread achieved by the current solution $S_{i,t}$ for the $i^{th}$ layer of $\mathscr{G}$. The solution set for the whole $\mathscr{G}$ would be:  

\begin{equation}
\label{eq:final_seed_whole_net}
    \hat{S} = \bigcup\limits_{i=1}^{k} S_{i}
\end{equation}





\subsection{MIM-Reasoner Analysis}



In order to bound the solution of MIM-Reasoner when policy $\pi$ is converged, it is necessary to ensure that the optimal policy $\pi^*$ follows a greedy strategy. The following lemma assists in ensuring this guarantee:

\begin{lemma}
(Greedy Policy Guarantee). When policy $\pi$ is converged, $\pi^*(v \mid S_{i,t})$ always selects nodes greedily at every time step $t$. (Proof in Appendix \textbf{B.2})

\label{lemma:optimal Greedy Policy Convergence}
\end{lemma}

Based on Lemma \ref{lemma:optimal Greedy Policy Convergence}, for every layer $G_{i} \in \mathscr{G}$, the solution provided by policy $\pi^*$ is within $(1-1/e)$ of the optimal solution, given that $\sigma_i$ is submodular and monotone increasing. Our next concern lies in the quality of 
MCKP. During phase 1, since 
$\mathcal{A}$ runs in parallel for each layer $G_{i}$, 
we also achieve an approximation ratio of $(1-1/e)$ as stated in the following lemma. 



\begin{lemma}
(Multiple Choice Knapsack Problem Guarantee). Let $Opt_{S}$ be the value of the solution for MCKP instance $S$, and $Opt_{\Tilde{S}}$ be the value of the optimal solution for $\Tilde{S}$, we have: $O p t_{S} \geq(1-1/e) O p t_{\Tilde{S}}$ (Proof in Appendix \textbf{B.3})

\label{lemma:Multiple Choice Knapsack Problem Guarantee}
\end{lemma}

To ensure the solution is bounded, it is important to include near-optimal or optimal nodes in the search space $\mathcal{I}(.)$. 
If $\mathcal{I}(.)$ consists of $K$ layers, the prediction of any node $v \in V$ is influenced by its $K$-hop neighbors (\cite{zhou2018graph}). Although the graph itself can be changed, the underlying model generating the graph often remains consistent. 
These observations lead to an assumption that poor nodes will be eliminated in the solution set, while only good nodes are retained by the GNN-based model \citep{manchanda2020gcomb}. Thus, it is possible that $\mathcal{I}(.)$ trained on a set of subgraphs $\mathcal{E}$ can also find solutions on similar graphs in $\mathscr{G}$. We formally restate these observations in the following assumption:


\begin{Assumption}
(Near Optimal Nodes Are Included In Search Space). In each layer $G_i \in \mathscr{G}$, the near optimal seed nodes $S^{*}$ are included in $V^g$ provided by pretrained $\mathcal{I}(.)$. In other word, we have $S^{*} \subseteq V^g$.

\label{asp:havingOptimalSolution}
\end{Assumption}


We now approximate the solution quality in the worst case. 
Let us denote $o$, $\hat{S}^{\pi^*}$ as the total number of overlapping nodes in $\mathscr{G}$ and final solution found by $\pi^*$. The optimal policy $\pi^*$ has an approximation guarantee for the worst case as follows:

\begin{thm}
    (Approximation Ratio In The Worst Case) Suppose the propagation $\sigma_i$ on each layer of the multiplex is submodular, the optimal policy $\pi^*$ will find a solution $\hat{S}$ for multiplex network $\mathscr{G}$ with an approximation ratio of $\frac{(1-\epsilon)(1-1/e)}{(o+1) k}$. (Proof in Appendix \textbf{B.4})
    
    \label{thm:MIM-Reasoner's solution quality in worst case}
\end{thm}

In case the optimal policy $\pi^*$ finds the solution $S^{\pi^*}_i$ for each layer $G_i$ that completely avoids reactivating nodes already activated by other layers, we can express the approximation ratio of $\pi^*$ for the best case. 

\begin{thm}
    (Approximation Ratio In The Best Case). Assume the $\pi^*$ can avoid reactivating all the activated nodes, the spread of solution given by optimal policy $\pi^*$ is at least: $\sigma\left(\hat{S}^{\pi^*}\right) \geq \frac{(1-\epsilon)(1-1/e)}{k+o} \sigma(\tilde{S})$
    
    \label{thm:MIM-Reasoner's solution quality in best case}
\end{thm}

Let us denote $\beta \in [0,1]$ as the percentage of nodes that cannot be successfully avoided reactivation by the policy $\pi^*$, we can establish performance guarantees in a typical scenario where the optimal policy $\pi^*$ can partially avoid reactivating nodes that have already been activated by other layers. 

\begin{thm}
    (Approximation Ratio In The General Case). Assume the $\pi^*$ can avoid to partially reactivate the activated nodes by other layers. Thus, with $\beta \in [0,1]$, the spread of solution given by optimal policy $\pi^*$ is at least: $\sigma\left(\hat{S}^{\pi^*}\right) \geq \frac{(1-\epsilon)(1-1/e)}{(k-1)\beta o+o+k} \sigma (\tilde{S})$
    \label{thm:MIM-Reasoner's solution quality in general case}
\end{thm}

Proofs of Theorem \ref{thm:MIM-Reasoner's solution quality in best case} and Theorem \ref{thm:MIM-Reasoner's solution quality in general case} are shown in Appendix \textbf{B.5-B.6}. The time complexity of the MIM-Reasoner depends on the total number of PGMs stored in $\mathcal{G}^{prev}$. We have the following lemma:

\begin{lemma}
    (PGMs's Time complexity). The time complexity of structure learning for $\mathcal{G}^{prev}$ after $k$ selection step is $|\mathcal{Y}|^{2} \cdot (k-1)$. (Proof in Appendix \textbf{B.7})
    
    \label{lemma:timecomplexpgm}
\end{lemma}

Given Lemma \ref{lemma:timecomplexpgm}, we now bound the time complexity of MIM-Reasoner. Assuming we have $k$ layers, the time complexity of algorithm $A$ on $l$ seed nodes with graph $G_h$ (representing the layer with the highest number of nodes and edges) is denoted as $\textit{ tc}\left(A, G_h, l\right)$. The time complexity of training MIM-Reasoner is described as:

\begin{thm}
    (Time complexity of MIM-Reasoner).  The time complexity of the Budget Allocation is: ${\underset{h\in k}{\max }} \textit{ tc}\left(A, G_h, l\right)+(k l)^{\lceil 1 / \epsilon-1\rceil} \log k$ and the time complexity of Relation RL Optimization is :
    $O(|\mathcal{Y}|^{2} \cdot (k-1) +\mathcal{Q})$ where $\mathcal{Q}$ is number of step for policy $\pi$ converge to optimal. (Proof in Appendix \textbf{B.8})
    
    \label{thm:timecomplexpgm}
\end{thm}

\begin{figure*}[t]
    \centering
    \includegraphics[width=1\linewidth]{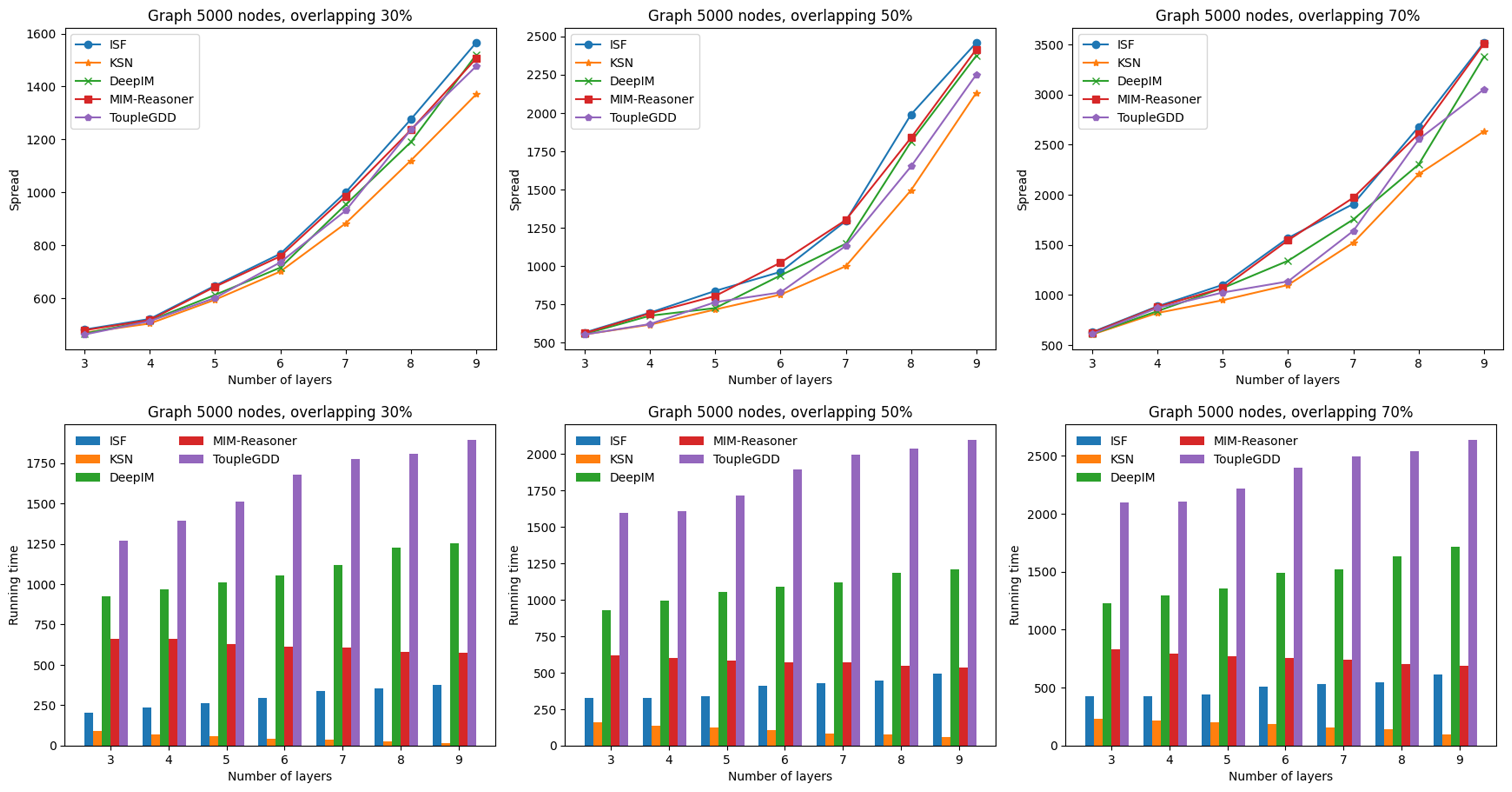}
    \caption{Comparison of five methods on a synthetic dataset, with different overlapping percentages and layers. The comparison is based on two metrics: total spreading and running time (in seconds).}
    \label{fig:synthetic}
\end{figure*}

\begin{table*}[ht]

\centering
\resizebox{\textwidth}{!}{
\begin{tabular}{c|ccc|ccc|ccc|ccc}
\toprule
\multirow{2}{*}{\textbf{Methods}} & \multicolumn{3}{c|}{\textbf{Celegans}}                                                                       & \multicolumn{3}{c|}{\textbf{Drosophila}}                                                                     & \multicolumn{3}{c|}{\textbf{Twitter-Foursquare}}                                                             & \multicolumn{3}{c}{\textbf{Pope-Election}}                                                                  \\ \cmidrule{2-13} 
                                  & \multicolumn{1}{c}{\textbf{TS}} & \multicolumn{1}{c}{\textbf{RT}} & \textbf{IT} & \multicolumn{1}{c}{\textbf{TS}} & \multicolumn{1}{c}{\textbf{RT}} & \textbf{IT} & \multicolumn{1}{c}{\textbf{TS}} & \multicolumn{1}{c}{\textbf{RT}} & \textbf{IT} & \multicolumn{1}{c}{\textbf{TS}} & \multicolumn{1}{c}{\textbf{RT}} & \textbf{IT} \\ \midrule
ISF                               & \multicolumn{1}{c}{\textbf{1412.21}}               & \multicolumn{1}{c}{335.96}                & \textbf{NaN}       & \multicolumn{1}{c}{3268.22}               & \multicolumn{1}{c}{3087.58}               & \textbf{NaN}       & \multicolumn{1}{c}{\textbf{x}}            & \multicolumn{1}{c}{\textbf{x}}            & \textbf{NaN}       & \multicolumn{1}{c}{\textbf{x}}            & \multicolumn{1}{c}{\textbf{x}}            & \textbf{NaN}       \\ 
KSN                               & \multicolumn{1}{c}{1267.14}               & \multicolumn{1}{c}{\textbf{176.02}}       & \textbf{NaN}       & \multicolumn{1}{c}{2911.46}               & \multicolumn{1}{c}{\textbf{1331.23}}      & \textbf{NaN}       & \multicolumn{1}{c}{49282.36}              & \multicolumn{1}{c}{\textbf{11820.24}}     & \textbf{NaN}       & \multicolumn{1}{c}{\textbf{x}}            & \multicolumn{1}{c}{\textbf{x}}            & \textbf{NaN}       \\ 
ToupleGDD                         & \multicolumn{1}{c}{1309.13}               & \multicolumn{1}{c}{1181.72}               & 1.12               & \multicolumn{1}{c}{3150.71}               & \multicolumn{1}{c}{10945.98}              & 2.74               & \multicolumn{1}{c}{49927.11}              & \multicolumn{1}{c}{71428.59}              & 7.46               & \multicolumn{1}{c}{246873.52}             & \multicolumn{1}{c}{317431.28}             & 13.21              \\ 
DeepIM                            & \multicolumn{1}{c}{1310.91}               & \multicolumn{1}{c}{715.11}                & \textbf{0.23}      & \multicolumn{1}{c}{3250.58}               & \multicolumn{1}{c}{8740.67}               & \textbf{0.49}      & \multicolumn{1}{c}{51989.23}              & \multicolumn{1}{c}{43158.84}              & \textbf{1.17}      & \multicolumn{1}{c}{248251.49}             & \multicolumn{1}{c}{251195.93}             & \textbf{1.98}      \\ \midrule
\textbf{MIM-Reasoner}             & \multicolumn{1}{c}{1314.25}      & \multicolumn{1}{c}{580.76}                & 0.28               & \multicolumn{1}{c}{\textbf{3462.94}}      & \multicolumn{1}{c}{4367.41}               & 0.57               & \multicolumn{1}{c}{\textbf{53420.94}}      & \multicolumn{1}{c}{28372.40}              & 1.81               & \multicolumn{1}{c}{\textbf{254238.20}}     & \multicolumn{1}{c}{\textbf{99713.08}}     & 2.27               \\ \bottomrule
\multicolumn{13}{r}
{TS: Total Spread; RT: Running Time (Seconds); IT: Inference Time (Seconds).}
\end{tabular}
}
\caption{Comparison of five methods on different real datasets in terms of total spreading, running time in seconds, and inference time in seconds. Best performance is highlighted with bold while \textbf{x} indicates an out-of-memory error. \textbf{NaN} means that traditional methods do not have an inference time. }\label{table:comparison}
\end{table*}

\section{EXPERIMENT EVALUATION}
This section compares the performance of MIM-Reasoner across four real multiplex networks and one synthetic multiplex network in maximizing the influence under various settings, following a case study to qualitatively demonstrate its performance. The code and datasets for MIM-Reasoner are available at the following GitHub repository: \url{https://github.com/nguyendohoangkhoiUF/MIM-Reasoner}.

\subsection{Experiment Setup}
Our purpose is to evaluate the expected influence spread as defined in Equation \eqref{eq: problem-definition} and time efficiency under a multiplex network scenario. For more detailed information about the experiment setup and our findings, we encourage readers to refer to Appendix \textbf{C.1 and C.2}.

\noindent\textbf{Comparison Methods And Metrics.} We compare the MIM-Reasoner with two sets of approaches. 1) Traditional multiplex influence maximization methods: \textit{ISF (Influential Seed Finder)} \citep{kuhnle2018multiplex} is a greedy algorithm designed for multiplex influence maximization; \textit{KSN (Knapsack Seeding of Networks)} \citep{kuhnle2018multiplex} utilizes a knapsack approach to find the best seed users in a multiplex network. 2) Deep Learning based influence maximization methods for single network: \textit{ToupleGDD} \citep{chen2022touplegdd} is a Deep Reinforcement Learning-based solution trained on many small networks for better generalization ability; \textit{DeepIM} \citep{ling2023deep} is a state-of-the-art IM solution based on deep generative models. The comparison is based on three metrics: total influence spread (activated nodes), running time (in seconds) and inference time (in seconds).

\noindent\textbf{Synthetic Dataset:} We compare MIM-Reasoner with other baselines on a synthetic multiplex network with 5,000 nodes and 25,000 edges in various overlapping rates ($30\%$, $50\%$, and $70\%$) and the number of layers (ranging from 3 to 9). 

\noindent\textbf{Real Dataset:} We evaluate MIM-Reasoner on four real-world multiplex networks: 1) Celegans \citep{celegans1}: 6 layers, 3879 nodes, and 8191 edges; 2) Drosophila \citep{celegans1,celegans2}: 7 layers, 8215 nodes, and 43,366 edges; 3) Twitter-Foursquare network~\citep{TFdataset}: 2 layers, 93269 nodes, and 17,969,114 edges; 4) Pope-Election \citep{De2020unraveling}: 3 layers, 2,064,866 nodes, and 5,969,189 edges.

\noindent\textbf{Hyperparameter Setting.}
The experiments were run with budget $l = 30$, and Monte Carlo simulation $mc = 100$. The early stopping was also applied for all methods. Specifically, DeepIM is stopped when the increment of reconstruction accuracy is not larger than $\epsilon$ for every 25\% of the total number of epochs. For ToupleGDD and MIM-Reasoners, for every 25\% of the total number of epochs, the early stop was applied if the current solution was not better than the best one.

\subsection{Training Time Analysis}
The ``running time'' of combinatorial algorithms (such as ISF or KSN) and learning-based approaches (such as DeepIM, ToupleGDD, and MIM-Reasoner) is interpreted as follows. Combinatorial algorithms measure running time as the duration to find a solution, while learning-based approaches refer to the running time as the time needed for training the model to converge and achieve a solution with a total spread reported. Therefore, this section focuses on analyzing the running time (training time) of learning-based methods. 

Compared with Deep Learning-based influence maximization solutions, i.e., DeepIM and ToupleGDD, the running time for MIM-Reasoner is typically the lowest in both synthetic and real dataset as depicted in Figure \ref{fig:synthetic} and Table \ref{table:comparison}. By parallelizing the IC model on each individual layer for every optimization step, MIM-Reasoner reduces the complexity of the propagation model significantly. In contrast, the size of the network directly affects the number of training samples needed for DeepIM, resulting in longer training times to provide decent solutions. Meanwhile, ToupleGDD running the propagation model on the entire multiplex network (with a much larger node and edge count) would naturally require more computational time for each step. 

Interestingly, as illustrated in Figure \ref{fig:synthetic}, the training time of MIM-Reasoner decreases as the number of layers increases. In fact, it becomes comparable to the running time of combinatorial algorithms such as ISF in the synthetic dataset. This demonstrates that MIM-Reasoner has an advantage of reducing training time as layer complexity increases, while still achieving competitive performance in terms of total spread.

\subsection{Inference Time Analysis}
As shown in Table \ref{table:comparison}, MIM-Reasoner stands out as an efficient method, particularly because it can parallelly process separate layers of a multiplex network and conduct batch inference. Even though DeepIM is the fastest because it predicts end-to-end solutions directly; however, the training of DeepIM entails learning a large model, which is resource-consuming. ToupleGDD has to infer step by step each seed node, which makes the inference time longer than others. Moreover, it is crucial to compare the running time of combinatorial algorithms with the inference time of learning-based approaches. For instance, even the fastest running combinatorial algorithm like KSN, which takes $176.02$s and $1331.23$s to produce a solution in two small real network datasets respectively, cannot be compared to the efficiency of MIM-Reasoner, which only requires 0.28s and 0.57s. When considering larger graph sizes such as Twitter-Foursquare and Pope-Election, MIM-Reasoner still performs impressively, taking only 1.81s and 2.27s respectively, while producing a comparable total spread. In contrast, KSN takes 11820.24s for Twitter-Foursquare and cannot provide a solution for Pope-Election. Furthermore, ISF fails to provide solutions for both the enormous network Twitter-Foursquare and Pope-Election. This comparison highlights the advantages of using machine learning-based approaches when addressing the MIM problem.

\subsection{Quantitative Analysis}
\noindent\textbf{Synthetic Dataset.} In terms of other comparison methods, ISF, KSN, and DeepIM show similar trends in spreading, while ToupleGDD typically lags behind, particularly with higher overlap percentages. Among all the approaches evaluated, MIM-Reasoner consistently demonstrates competitive spreading values across all three overlapping percentages. It outperforms other methods, particularly as the number of layers in the multiplex network increases. Even with small graphs created in the synthetic dataset, MIM-Reasoner provides a total spread that is nearly as good as that of ISF. It is worth noting that while MIM-Reasoner may trade off solution quality for faster training and inference times, it still provides solutions with a total spread comparable to ISF. This is significant because ISF has a good approximation ratio $(1-1/e)$ in multiplex networks.


\noindent\textbf{Real-world Dataset.} The results of MIM-Reasoner on real-world datasets are summarized in Table \ref{table:comparison}. Among all the approaches evaluated, MIM-Reasoner consistently achieves either the highest or near-highest spreading across all datasets while providing fast inference times to generate solutions. This suggests the effectiveness of MIM-Reasoner in maximizing influence. In contrast, the performance of other methods, including ISF with its $1-1/e$ ratio in multiplex networks, varies across datasets, with none consistently outperforming MIM-Reasoner in terms of spreading. Traditional methods, such as ISF and KSN, can provide solutions with total spreads comparable to that of MIM-Reasoner. However, their running times are significantly slower than the inference time of MIM-Reasoner. Moreover, these traditional methods encounter issues such as out-of-memory errors or excessively long running times (e.g., running for over two weeks) on larger datasets like "Twitter-Foursquare" and "Pope-Election". This indicates their poor scalability in real-world scenarios.

\section{CONCLUSION}
We propose a novel framework named MIM-Reasoner to tackle the multiplex influence maximization problem. Specifically, we leverage deep reinforcement learning and decompose the multiplex network into separate layers. Our framework learns to allocate node selection budgets for each layer and employs a lightweight policy to find feasible solutions between each layer. A probabilistic graphical model is then designed to capture the propagation pattern between each layer in order to maximize the overall spread and improve the overall efficiency. The approximation ratio of the solution is provided with a theoretical guarantee. Finally, the proposed framework is compared with several SOTA approaches and demonstrates overall competitive performance on four real-world datasets and a synthetic dataset.

\section*{Acknowledgements}

This material is partially supported by the National Science Foundation (NSF) under Grant No. FAI-1939725, SCH-2123809, and IIS-1908594, and the Department of Homeland Security (DHS) under Grant No. 17STCIN00001.

\bibliography{main}




\section*{Checklist}



 \begin{enumerate}

 \item For all models and algorithms presented, check if you include:
 \begin{enumerate}
   \item A clear description of the mathematical setting, assumptions, algorithm, and/or model. [Yes]
   \item An analysis of the properties and complexity (time, space, sample size) of any algorithm. [Yes]
   \item (Optional) Anonymized source code, with specification of all dependencies, including external libraries. [Yes]
 \end{enumerate}

 \item For any theoretical claim, check if you include:
 \begin{enumerate}
   \item Statements of the full set of assumptions of all theoretical results. [Yes]
   \item Complete proofs of all theoretical results. [Yes]
   \item Clear explanations of any assumptions. [Yes]     
 \end{enumerate}

 \item For all figures and tables that present empirical results, check if you include:
 \begin{enumerate}
   \item The code, data, and instructions needed to reproduce the main experimental results (either in the supplemental material or as a URL). [Yes]
   \item All the training details (e.g., data splits, hyperparameters, how they were chosen). [Yes]
    \item A clear definition of the specific measure or statistics and error bars (e.g., with respect to the random seed after running experiments multiple times). [Yes, except for error statistics in the tables. The reason is the space limit.]
    \item A description of the computing infrastructure used. (e.g., type of GPUs, internal cluster, or cloud provider). [Yes]
 \end{enumerate}

 \item If you are using existing assets (e.g., code, data, models) or curating/releasing new assets, check if you include:
 \begin{enumerate}
   \item Citations of the creator If your work uses existing assets. [Yes]
   \item The license information of the assets, if applicable. [Yes]
   \item New assets either in the supplemental material or as a URL, if applicable. [Yes]
   \item Information about consent from data providers/curators. [Yes]
   \item Discussion of sensible content if applicable, e.g., personally identifiable information or offensive content. [Yes]
 \end{enumerate}

 \item If you used crowdsourcing or conducted research with human subjects, check if you include:
 \begin{enumerate}
   \item The full text of instructions given to participants and screenshots. [Not Applicable]
   \item Descriptions of potential participant risks, with links to Institutional Review Board (IRB) approvals if applicable. [Not Applicable]
   \item The estimated hourly wage paid to participants and the total amount spent on participant compensation. [Not Applicable]
 \end{enumerate}

 \end{enumerate}


\clearpage
\onecolumn
\clearpage
\onecolumn
\appendix
\onecolumn

\aistatstitle{MIM-Reasoner: Learning with Theoretical Guarantees for Multiplex
Influence Maximization}

\textbf{A. Detail Steps of MIM-Reasoner}

\begin{algorithm}
\DontPrintSemicolon
\SetKwInput{KwInput}{Input}
\SetKwInput{KwOutput}{Output}
\SetKw{KwTo}{to}
\SetKw{KwFrom}{from}
\SetKw{KwWith}{with}
\SetKw{KwCalculate}{calculate}
\SetKw{KwStore}{store}
\SetKw{KwReturn}{return}
\SetKwFunction{Sort}{sort}
\SetKwFunction{RunAlgorithm}{runAlgorithm}
\SetKwFunction{ChooseSeedSets}{chooseSeedSets}

\KwInput{Algorithm $\mathcal{A}$, a multiplex network $\mathscr{G} = (G^1, G^2, \ldots, G^k)$, budget $l$}
\KwOutput{Budget allocation table $\mathcal{U}$}

Initialize $\mathcal{S} := \emptyset$

\ForEach{$G^i \in \mathscr{G}$}{
$S_{ij} :=$ \RunAlgorithm{$\mathcal{A}$, $G^i$, $l$} \tcp*{Run any algorithm A on each layer}
$\mathcal{S} := \mathcal{S} \cup S_{ij}, \forall j \in [1,...,l] $ \tcp*{Collect found solution}
}

$C := \emptyset; P := \emptyset$

\ForEach{$S_{ij} \in \mathcal{S}$}{
$p := \sigma(S_{ij}); c := |S_{ij}|$ \tcp*{Calculate the profit and cost}

$P := P \cup [p]; C := C \cup [c]$ \tcp*{Store the profit and cost}
}

$\mathcal{H} := \mathcal{H}[P][C]$ \tcp*{Create a Profit-Cost table with profits $P$ as rows and costs $C$ as columns}

$\mathcal{U} := \text{MCKP-Solver}(\mathcal{H})$ \tcp*{Solve the MCKP to obtain budget allocation $\mathcal{U}$}

\KwReturn{$\mathcal{U}$};

\caption{Budget Allocation (Phase 1)}
\label{alg:budget-allocation}
\end{algorithm}

\begin{algorithm}
\DontPrintSemicolon
\SetKwInput{KwInput}{Input}
\SetKwInput{KwOutput}{Output}
\SetKw{KwTo}{to}
\SetKw{KwFrom}{from}
\SetKw{KwWith}{with}
\SetKw{KwCalculate}{calculate}
\SetKw{KwStore}{store}
\SetKw{KwReturn}{return}
\SetKwFunction{Sort}{sort}

\KwInput{Budget Allocation table $\mathcal{U}$}
\KwOutput{RL model $\pi$}

Initialize $\mathcal{G}^{prev} := \emptyset$, $\mathcal{D} := \emptyset$

\For{$i \in {1, \ldots, k}$}{
Select layer $G_i = \arg\min_{q \in {1,...,k}} \mathcal{U}[q]$ \tcp*[r]{Select layer with minimum budget}

Remove row $i^{th}$ in table $\mathcal{U}$ \tcp*[r]{Prevent reselection}

$j:=U[i][0]$ \tcp*[r]{Select budget for layer $G_i$}

\If{$\mathcal{D} = \emptyset$}{
    Collect seed set $S_{ij}$ from phase 1 \tcp*[r]{Collect seed set from phase 1}
    Train policy $\pi$ to learn seed set $S_{ij}$ and record $\mathcal{D}$ \tcp*[r]{Train policy with seed set}
}
\Else{
    Train $\pi$ with budget $j$ based on $\mathcal{G}^{prev}$ and record a new $\mathcal{D}$ \tcp*[r]{Train policy with rewards generated from $\mathcal{G}^{prev}$}
}

$\mathcal{Y} := \operatorname{Variable Grouping}(\mathcal{D})$ \tcp*[r]{Perform variable grouping}
$\varphi_{i} := \operatorname{Structure Learning}(\mathcal{Y})$ \tcp*[r]{Training PGM}
$\mathcal{G}^{prev} := \mathcal{G}^{prev} \cup \varphi_{i}$ \tcp*[r]{Update previous structure set}

}

\KwReturn{$\pi$}

\caption{Relation RL Optimization (Phase 2)}
\label{alg:Relation RL Optimization}
\end{algorithm}

\textbf{A.1. Good Candidate Finding}






\begin{algorithm}
\DontPrintSemicolon
\SetKwInput{KwInput}{Input}
\SetKwInput{KwOutput}{Output}
\SetKw{KwTo}{to}
\SetKw{KwFrom}{from}
\SetKw{KwWith}{with}
\SetKw{KwCalculate}{calculate}
\SetKw{KwStore}{store}
\SetKw{KwReturn}{return}
\SetKwFunction{Sort}{sort}
\SetKwFunction{RunAlgorithm}{runAlgorithm}
\SetKwFunction{ChooseSeedSets}{chooseSeedSets}

\KwInput{Graph $\hat{G}_{i,z}=(V_z,E_z)$, propagation function $\sigma_i(.)$ of layer $G_i$, convergence threshold $\Delta$}
\KwOutput{Solution set $S^q_z$, $|S^q_z|=b$}

$S^q_z \leftarrow \emptyset$ \tcp*{Initialize solution set}

\While{ $\digamma > \Delta$}{
$v \leftarrow$ Choose with probability $\frac{\sigma_i(S^q_z \cup{v})-\sigma_i(S^q_z)}{\sum_{\forall v^{\prime} \in V \backslash S} \sigma_i\left(S^q_z \cup\{v^{\prime}\}\right)-\sigma_i(S^q_z)}$ \tcp*{Choose vertex probabilistically}
$\digamma \leftarrow f(S^q_z \cup{v})-f(S^q_z)$ \tcp*{Calculate gain}
$S^q_z \leftarrow S^q_z \cup{v}$ \tcp*{Add selected vertex to the solution set}
}

\KwReturn{$S^q_z$} \tcp*{Return solution set}

\caption{The Probabilistic Greedy Approach}
\label{alg:probabilistic-greedy}
\end{algorithm}

Even when dividing the multiplex network into separate layers, each layer $G_i \in \mathscr{G}$ can still be at a billion-scale. The objective of Good Candidate Finding is to identify nodes that are unlikely to be part of the solution set, thereby reducing the search space. This is possible because although the graph itself can change, the underlying model generating the graph often remains consistent. Let $\mathcal{E}_i = [\hat{G}_{i,1},...,\hat{G}_{i,h}]$ be a set of subgraphs induced from the $i^{th}$ layer of $\mathscr{G}$, where $h$ is a hyperparameter representing the total number of subgraphs the trainer wants to sample. For each layer $G_i \in \mathscr{G}$, we sample a set of subgraphs $\mathcal{E}_i$, hence, our graph training dataset becomes $\mathcal{E} = [\mathcal{E}_1, \mathcal{E}_2, \cdots, \mathcal{E}_k]$. We utilize a classification-based method to train the Graph Attention Network (GAT) model, denoted as $\mathcal{I}$. In this method, given any training graph $\hat{G}_{i,z} \in \mathcal{E}_i$, a budget $b$, and its corresponding solution set $S_{z}$, a node $v$ is labeled as positive if $v \in S_z$, and negative otherwise. 

In certain cases, within each graph $\hat{G}_{i,z}=(V_z,E_z)$, there can be situations where two nodes $(v_1, v_2) \in V_z$ have the same influence spread (i.e., $\sigma_i(v_1) = \sigma_i(v_2)$). However, when considering the marginal gain of adding node $v_2$ to the solution set $S_z = {v_1}$, denoted as $\sigma_i((v_1, v_2)) - \sigma_i((v_1))$, it turns out to be zero and vice versa. Consequently, in such scenarios, even though both nodes have equal and individual quality, only one of them would be selected in the final answer set meaning only one of the two nodes is considered as a positive candidate. That is the reason why we need to use a Probabilistic Greedy strategy to ensure that both $v_1$ and $v_2$ are considered positive candidates (good candidates).

\textbf{Probabilistic greedy}: For each graph $\hat{G}_{i,z}\in \mathcal{E}_i$, we employ a probabilistic greedy approach that involves sampling a node from the search space $V_z$ while considering marginal gain of the sampled node. Specifically, instead of always selecting the node $v \in V_z$ with the highest marginal gain, we choose $v$ in a probabilistic manner where $v$ is chosen with a probability that is proportional to its marginal gain (Algorithm \ref{alg:probabilistic-greedy}). During each iteration, the probabilistic greedy algorithm performs $\varkappa$ times to create $\varkappa$ different solution sets $\mathbb{S}_{z}=\{S^1_{z}, \cdots, S^{\varkappa}_z\}$. To determine the score of a node $v \in V_z$, the algorithm assigns a value based on the following process:

\begin{equation}
\operatorname{\omega}(v)=\frac{\sum_{q=1}^{\varkappa} \digamma_q(v)}{\sum_{q=1}^{\varkappa} \sigma_i\left(S_{z}^q\right)}    
\end{equation}

Here, $S_{z}^q$ is the $q^{th}$ seed set of graph $\hat{G}_{i,z}$ while  $\digamma_q(v)$ is the marginal gain contribution of $v$ to $S_z^q$. Suppose we have a seed set $S_{z}^q=\{1,3,5,4\}$, the marginal gain contribution of node $v=3$ can be computed as $\digamma_q(3)=\sigma_i((1, 3)) - \sigma_i((1))$. Based on the set $\mathbb{S}_{z}$, we can have a set of good nodes $V^g_z=\bigcup_{q=1}^{\varkappa} S_z^q$. For each good node $v \in V^g_z$, we want $\mathcal{I}$ with parameters $\Theta$ predict $\hat{\omega}(v)$ as close as much as possible with the found $\omega(v)$. Therefore, we use Mean Square Error (MSE) as the objective function as shown here:
\begin{equation}
J\left(\Theta\right)=\sum_{\sim\left\langle G_{i,z}\right\rangle} \frac{1}{\left|V_z^g\right|} \sum_{\forall v \in V_z^g}[\operatorname{\omega}(v)-\hat{\omega}(v)]^2    
\end{equation}

In the above equation, $V_z^g$ denotes the set of good nodes in graph $G_i$. Since $\mathcal{I}$ is trained through message passing, in a $\mathcal{I}$ with $K$ hidden layers, the computation of each node $v$ is limited to the induced subgraph formed by its $K$-hop neighbors, instead of the entire graph.

\textbf{A.2. Variables Grouping}
\begin{algorithm}
\DontPrintSemicolon
\SetKwInput{KwInput}{Input}
\SetKwInput{KwOutput}{Output}
\SetKw{KwTo}{to}
\SetKw{KwFrom}{from}
\SetKw{KwWith}{with}
\SetKw{KwCalculate}{calculate}
\SetKw{KwStore}{store}
\SetKw{KwReturn}{return}
\SetKwFunction{Sort}{sort}
\SetKwFunction{RunAlgorithm}{runAlgorithm}
\SetKwFunction{ChooseSeedSets}{chooseSeedSets}

\KwInput{Graph $\hat{G}_{i,z}=(V_z,E_z)$, Status dataset $\mathcal{D}$, correlation threshold $\xi$}
\KwOutput{Set of representative nodes $\mathcal{Y}$}

$\mathcal{P}:=\emptyset$; \tcp*{Initialize an empty set of variable groups}

$\mathcal{Y}:=\emptyset$; \tcp*{Initialize an empty set of representative nodes}

$Q=\operatorname{PearsonCorrelation}(\mathcal{D})$ \tcp*{Calculate Pearson Correlation for all node in $\mathcal{D}$}

Initialize an array $isolated$ having $|V_z|$ elements, and set each element to True\; \tcp*{To check whether a node has any group}

$S=\operatorname{NaN-Grouping}(Q)$ \tcp*{Group nodes with constant status together to create a new set of nodes with changing status}

\ForEach{$x \in S$}{
\If{isolated[x]=True}{ 
Create a new group $P_i \in \mathcal{P}$;

$P_i := P_i \cup x$; \tcp*{Add $x$ to $P_i$}

$\mathcal{P}:= \mathcal{P} \cup P_i$; \tcp*{Add $P_i$ to $\mathbf{P}$}

    \ForEach{$y \in S$}{
        \If{isolated[y]=True and $Q[x, y] > \xi$ and $Q[x, y] \neq "NaN"$}{
            $P_i:=P_i \cup y$; \tcp*{Add $y$ to $P_i$ }
        }
    }
    
    Find the representative node $v_i$ in $P_i$ by selecting the node closest to the centroid of $P_i$\;
    $\mathcal{Y} := \mathcal{Y} \cup v_i$\; \tcp*{Add representative node $v_i$ for group $P_i$ to $\mathcal{Y}$}
}
}

\KwReturn{$\mathcal{Y}$};

\caption{Variable Grouping}
\label{alg:variable-grouping}
\end{algorithm}




Recall the fact that the number of nodes in the Status dataset $\mathcal{D}$ can be at billion-scale, it is necessary to reduce complexity before training a Probabilistic Graphical Model through Structure Learning. One preprocessing step that achieves this is Variable Grouping. Variable Grouping involves grouping highly correlated nodes in the Status dataset $\mathcal{D}=[0,1]^{m \times |V|}$ together, resulting in a smaller set $\mathcal{P}=\left[P_1, P_2, \ldots, P_q\right]$ where $q$ is the total number of groups. This grouping satisfies the property that $V=\bigcup_{i=1}^q P_i$. The underlying idea is that when nodes are highly correlated, they often share similar characteristics or behaviors. For example, let's consider two nodes $(v_1,v_2)$ in dataset $\mathcal{D}$. In a Monte Carlo Simulation step with $m$ iterations, it is observed that when $v_1$ is activated, there is a 70 $\%$ percentage that $v_2$ is also activated. This means that $v_1$ and $v_2$ can have high correlations and similar patterns of activation. By knowing only a representative node, we can infer the properties of other nodes in the cluster without explicitly including them in the PGM, hence, reducing the complexity. In this work, we use Pearson Correlation to measure the correlation between two variables.

\textbf{Pearson Correlation metric:} Pearson correlation is a statistical measure that quantifies the linear relationship between two variables. It is used to assess the strength and direction of the relationship between two variables and it is denoted as $\chi \in [-1,1]$. A positive value of $\chi$ indicates a positive linear relationship, meaning that as one variable increases, the other tends to increase as well. Meanwhile, a negative value of $\chi$ indicates a negative linear relationship, where as one variable increases, the other tends to decrease. Lastly, $\chi=0$ indicates no linear relationship between the variables. Given two variables $x$ and $y$ and paired data $\{\left(x_1, y_1\right), \ldots,\left(x_m, y_m\right)\}$ consisting of $m$ pairs, the Pearson correlation coefficient is computed as follows:

\begin{equation}
\chi_{xy} = \frac{\sum_{i=1}^m\left(x_i-\bar{x}\right)\left(y_i-\bar{y}\right)}{\sqrt{\sum_{i=1}^m\left(x_i-\bar{x}\right)^2} \sqrt{\sum_{i=1}^m\left(y_i-\bar{y}\right)^2}}    
\end{equation}

Here, $m$ represents the sample size, $x_i$ and $y_i$ represent the individual sample points indexed with $i$, $\bar{x}=\frac{1}{m} \sum_{i=1}^m x_i$ represents the sample mean of $x$, and $\bar{y}$ represents the sample mean of $y$. 

We base on the Pearson correlation coefficient defined above to calculate a correlation coefficient for all pairs of variables (nodes) in dataset $\mathcal{D}$ to obtain the correlation matrix $Q^{|V| \times |V|}$ where $Q_{x,y} = \chi_{xy} | 1 \leq x, y \leq |V|$. We then define a threshold $\xi$ and traverse the correlation matrix $Q$. Any two variables with a correlation coefficient higher than $\xi$ are considered highly correlated and should be grouped (Algorithm 2). An important thing to note is that there will be nodes that never change status, such as always active nodes or always de-active nodes. In that case, the correlation of those nodes to any remaining nodes will become ``NaN'' because their standard deviation (the denominator in $\chi_{xy}$ equation) becomes zero. Those nodes will be grouped with other nodes $v \in V$. After grouping nodes in dataset $\mathcal{D}$, we will have a set of highly correlated node groups $\mathcal{P}=\left[P_1, P_2, \ldots, P_q\right]$ . The next step is to define a representative node for each group $P_i \in \mathcal{P}$. In this work, we take the node closest to the centroid from each group as a representative node. The set of representative nodes is denoted as $\mathcal{Y}=\left[v_1, \ldots, v_q\right]$, where each representative node $v_i \in \mathcal{Y}$ corresponds to a group $P_i \in \mathcal{P}$.


\textbf{A.3. Structure Learning}






Structure learning in Probabilistic Graphical Models (PGMs) involves automatically discovering the optimal graphical structure that represents the relationships between random variables in a dataset. It helps infer conditional independence relationships, identify direct and causal dependencies, and understand the underlying causal mechanisms. This is crucial for tasks like prediction, inference, and gaining insights into the data.


After the Variables Grouping step, we have an updated set of representative variables $\mathcal{Y}=\left[v_1, \ldots, v_q\right]$, status dataset $\mathcal{D}$ which will be used to generate the structure of our Bayesian model. In this work, we employ the Tree-based algorithm called Chow Liu to learn structure because it offers a balance between complexity and representation power. Given any random directed tree, an approximation of the true joint probability distribution of $\mathcal{Y}$ is in the form as follows:
\begin{equation}
P_t(\mathcal{Y})=\prod_{i=1}^q P\left(v_i \mid v_{\Gamma(i)}\right)
\end{equation}

Here, $\Gamma(i)$ is the parent of node $v_i$. If a vertex $i$ is the root, its parent $\Gamma(i) = \emptyset$, and the conditional probability $P\left(v_i \mid v_{\Gamma(i)}\right)$ simplifies to $P(v_i)$. It is important to note that in a directed tree, we must first select a vertex as the root. The edges are oriented away from the root, resulting in each vertex having at most one parent (but potentially multiple children).

\textbf{Problem Definition:} Let $P(\mathcal{Y})$ be the true joint probability distribution of $\mathcal{Y}$. Given a set of representative variables $\mathcal{Y}=\left[v_1, \ldots, v_q\right]$ and the set of all possible first-order dependence trees denoted as $T_q$, we want to find the optimal first-order dependence tree or the optimal structure denoted as $\Psi$ such as $\mathrm{KL}\left(P, P_\Psi\right) \leq \mathrm{KL}\left(P, P_t\right)$ for all $t \in T_q$.


Chow Liu algorithm solves this Minimization Problem by searching a maximum weight spanning tree (MWST). Lets define the mutual information $I\left(v_i, v_j\right)=\sum_{v_i, v_j} P\left(v_i, v_j\right) \log \left(\frac{P\left(v_i, v_j\right)}{P\left(v_i\right) P\left(v_j\right)}\right)$ between two variables $v_i$ and $v_j$. The key insight of MWST is that a probability distribution of tree dependence $P_t(\mathcal{Y})$ is an optimum approximation to $P(\mathcal{Y})$ if its tree model has maximum branch weight $\sum_{i=1}^q I\left(v_i, v_{\Gamma(i)}\right)$. Mathematically, we have: 

\begin{equation}
\begin{aligned}
& \mathrm{KL}\left(P, P_t\right)=\sum_{\mathcal{Y}} P(\mathcal{Y}) \log P(\mathcal{Y})-\sum_{\mathcal{Y}} P(\mathcal{Y}) \sum_{i=1}^q \log P\left(v_i \mid v_{\Gamma(i)}\right) \\
& =\sum_{\mathcal{Y}} P(\mathcal{Y}) \log P(\mathcal{Y})-\sum_{\mathcal{Y}} P(\mathcal{Y}) \sum_{i=1, \neq \text { root }}^q \log \frac{P\left(v_i, v_{\Gamma(i)}\right)}{P\left(v_{\Gamma(i)}\right)} \\
& =\sum_{\mathcal{Y}} P(\mathcal{Y}) \log P(\mathcal{Y})-\sum_{\mathcal{Y}} P(\mathcal{Y}) \sum_{i=1, \neq \text { root }}^q \log \frac{P\left(v_i,v_{\Gamma(i)}\right)}{P\left(v_i\right) P\left(v_{\Gamma(i)}\right)}-\sum_{\mathcal{Y}} P(\mathcal{Y}) \sum_{i=1}^q \log P\left(v_i\right)
\end{aligned}
\label{eqn:klp,p_t}
\end{equation}

Moreover, we should note that $-\sum_\mathcal{Y} P(\mathcal{Y}) \log P\left(v_i\right)=-\sum_{v_i} P\left(v_i\right) \log P\left(v_i\right)$. To understand why we have this, suppose $\mathcal{Y}=\left(v_1, v_2\right)$, let all variables are binary, let $i=1$

\begin{equation}
\begin{aligned}
- & \sum_\mathcal{Y} P(\mathcal{Y}) \log P\left(v_i\right) \\
= & -\left[P\left(v_1=0, v_2=0\right) \log P\left(v_1=0\right)+P\left(v_1=0, v_2=1\right) \log P\left(v_1=0\right)+\right. \\
& \left.P\left(v_1=1, v_2=0\right) \log P\left(v_1=1\right)+P\left(v_1=1, v_2=1\right) \log P\left(v_1=1\right)\right] \\
= & -\left[P\left(v_1=0\right) \log P\left(v_1=0\right)+P\left(v_1=1\right) \log P\left(v_1=1\right)\right] \\
= & -\sum_{v_1} P\left(v_1\right) \log P\left(v_1\right)=-\sum_{v_i} P\left(v_i\right) \log P\left(v_i\right)
\end{aligned}    
\end{equation}

We can also see that $-\sum_{\mathcal{Y}} P(\mathcal{Y}) \log P(\mathcal{Y})$, $-\sum_{v_i} P\left(v_i\right) \log P\left(v_i\right)$ are entropy terms $H(\mathcal{Y})$, $H\left(v_i\right)$, respectively. Then, we can rewrite equation \ref{eqn:klp,p_t} as:

$$\mathrm{KL}\left(P, P_t\right)=-\sum_{\mathcal{Y}} P(\mathcal{Y}) \sum_{i=1, \neq \text { root }}^q \log \frac{P\left(v_i,v_{\Gamma(i)}\right)}{P\left(v_i\right) P\left(v_{\Gamma(i)}\right)}+\sum_{i=1}^q H\left(v_i\right)-H(\mathcal{Y})$$

Moreover, we also have:

\begin{equation}
\begin{aligned}
\sum_\mathcal{Y} P(\mathcal{Y}) \log \frac{P\left(v_i, v_{\Gamma(i)}\right)}{P\left(v_i\right) P\left(v_{\Gamma(i)}\right)} 
&=\sum_{v_i, v_{\Gamma(i)}} P\left(v_i, v_{\Gamma(i)}\right) \log \frac{P\left(v_i, v_{\Gamma(i)}\right)}{P\left(v_i\right) P\left(v_{\Gamma(i)}\right)} \\
&=I\left(v_i, v_{\Gamma(i)}\right)
\end{aligned}
\label{eqn:I(v_i,u_parent)}
\end{equation}

Replace this result of equation \ref{eqn:I(v_i,u_parent)} to equation \ref{eqn:klp,p_t}, we have: 

\begin{equation}
\mathrm{KL}\left(P, P_t\right)=-\sum_{i=1}^q \underbrace{I\left(v_i, v_{\Gamma(i)}\right)}_{\begin{array}{c}
\text { Mutual information }
\end{array}}+\underbrace{\sum_{i=1}^q H\left(v_i\right)-H(\mathcal{Y})}_{\begin{array}{c}
\text { Independent of the } \\
\text { dependence tree }
\end{array}}
\end{equation}

Minimizing $\mathrm{KL}\left(P, P_t\right)$ is the same as maximizing the total branch weight $
\sum_{i=1}^n I\left(v_i, v_{\Gamma(i)}\right)
$. With this insight, Chow Liu uses Kruskal’s algorithm to construct a maximum
weight-spanning tree as a structure for PGM.

\textbf{B. Proofs and Supplementary Lemmas
}

\textbf{B.1. Preliminaries}

\textbf{Definition 3} \textit{(Submodular)}.
For all $A, B \subseteq V_i$, we have:

\begin{equation}
\sigma_i(A)+\sigma_i(B) \geq \sigma_i(A \cup B)+\sigma_i(A \cap B)
\end{equation}


\textbf{Definition 4} \textit{(Monotone Increasing)}. An objective function $\sigma(S)$ is monotone increasing if 

\begin{equation}
\sigma(S) \leq \sigma(T), S \subset T
\end{equation}

\textbf{Definition 5} (Diminishing Returns). An objective function $\sigma(S)$ is diminishing return if $\sigma(S \cup u)-\sigma(S) \geq \sigma(T \cup u)-\sigma(T)$, $\forall u \in T$ and $S \subset T$.

\textbf{Definition 6} \textit{(Generalized Deterministic Submodular)}. Let $\sigma$ be a model of influence propagation on the multiplex. $\sigma$ satisfies the Generalized Deterministic Submodular Property (GDS) if the expected number of activations, given a seed set $S$, can be expressed as:
\begin{equation}
    \sigma(S)=\sum_{i=1}^k p_i \sigma_i(S),
\end{equation}
where each $\sigma_i$ with $i \in\{1, \ldots, k\}$ is a deterministic submodular model of influence propagation, and $p_i \in[0,1], \sum_{i=1}^k p_i=1$.

\textbf{Definition 7} (Multiple-choice knapsack problem (MCKP)). Let $(\mathscr{C}, k, l, c, p, B)$ be given. Here a set of $k$ classes, $\mathscr{C}={C_1, \ldots, C_k}$, where each class $C_i$ consists of $l$ objects denoted as $C_i={x_{i1}, \ldots, x_{il}}$, and a budget $B \geq 0$ the goal is to select one item, $x_i^{\prime}$, from each class in a way that maximizes the total profit, $\sum_{i=1}^k p(x_i^{\prime})$ while ensuring that the total cost, $\sum_{i=1}^k c(x_i^{\prime})$, does not exceed the budget constraint $\sum_{i=1}^k c(x_i^{\prime}) < B$. Here, $c$ and $p$ represent the cost and profit functions associated with the objects $x_{ij}$, respectively.

\textbf{B.2. Greedy Policy Convergence}

\textbf{Lemma 1 }(Greedy Policy Convergence). When policy $\pi$ is converged to optimal, $\pi^*(v \mid S_t)$ always selects nodes greedily at every time step $t$. 



\textbf{Proof.} Recall the fact that our reward function to train policy $\pi$ is computed by considering the marginal gain of adding node $v_n$ to the partially computed solution $S_t$, as follows:

\begin{equation}
r_t=\mathcal{M}\left(S_{i, t} \cup v_n\right)-\mathcal{M}\left(S_{i, t}\right)
\end{equation}

Here $\mathcal{M}$ is customized evaluation
function which measures the total spread in layer
$G_i$, given solution $S_i$. When policy $\pi$ parameterized by $\theta$ is converged to optimal, the optimal policy $\pi^*$ will follow these parameters:
\begin{equation}
\theta^*=\arg \max _\theta \sum_{t=1}^\mathcal{Q} \mathbb{E}_{\left(S_{i,t}, v_n\right) \sim p\left(S_{i,t}, v_n \mid \theta\right)}\left[\mathcal{M}\left(S_{i, t} \cup v_n\right)-\mathcal{M}\left(S_{i, t}\right)\right]
\end{equation}

Considering these optimal parameters, we can use contradiction to prove this lemma. For any time step $t$ progresses from 1 to $\mathcal{Q}$, if $\theta^*$ is truly optimal, then it must ensure that at every step $t$, the agent is making a decision to maximize expected immediate reward, which is to pick the node providing the highest marginal gain. If there was a better node $v^{\prime}_n$ that the agent didn't select at some step $t$, then $\mathbb{E}_{\left(S_{i,t}, v_n\right) \sim p\left(S_{i,t}, v_n \mid \theta\right)}\left[\mathcal{M}\left(S_{i, t} \cup v_n\right)-\mathcal{M}\left(S_{i, t}\right)\right]$ wouldn't be maximized, contradicting our assumption that $\theta^*$ is optimal. Hence, this proves the lemma and we conclude that the optimal policy $\pi^*\left(v \mid S_t\right)$ will select nodes greedily for all $t$.


\textbf{B.3. Multiple Choice Knapsack Problem Guarantee}

\textbf{Lemma 2}.(Multiple Choice Knapsack Problem Guarantee). Let $O p t_{S_{ij}}$ be the value of the solution for MCKP instance $S_{ij}$, and $O p t_{\tilde{S}_{ij}}$ be the value of the optimal solution for $\tilde{S}_{ij}$, we have: 

\begin{equation}
O p t_{S_{ij}} \geq (1-1 / e) O p t_{\tilde{S}_{ij}}
\end{equation}

\textbf{Proof.} Since in Phase 1, we decompose MIM into separate layers, find a solution for each layer, then, combine back using MCKP. We need to examine how the MCKP problem can be represented for the MIM problem. From there, we can determine the approximation ratio of the solution combined with MCKP in the context of MIM.

Given a MIM instance $(\mathscr{G}, k, l, \sigma)$, where $\mathscr{G}$ represents the graph, $k$ is the number of seed sets, and $l$ is the size of each seed set. For each pair $(i, j)$, where $1 \leq i \leq k$ and $1 \leq j \leq l$, let $\tilde{S}_{i j}$ be an unknown optimal seed set for $G_i$ that satisfies two conditions: $\tilde{S}_{i j} \subset V$ (subset of nodes in $G_i$) and $|\tilde{S}_{i j}| = j$ (size of the seed set). In addtition, let $S_{ij}$ found by an algorithm $\mathcal{A}$ be an approximation to $\tilde{S}_{i j}$ and $S_{i j} \subset V$. Then, based on \textbf{Definition 7}, $C_i=\left\{S_{i 0}, \ldots, S_{i l}\right\}, C_i^{o p t}=\left\{\tilde{S}_{i 0}^{o p t}, \ldots, \tilde{S}_{i l}^{o p t}\right\}$. Finally, let $\mathscr{C}=\left\{C_i: 1 \leq i \leq k\right\}, \mathscr{C}^{\text {opt }}=\left\{C_i^{o p t}: 1 \leq i \leq k\right\}$, and for each $i, j$, define $c\left(S_{i j}\right)=j, p\left(T_{i j}\right)=\sigma\left(S_{i j}\right)$, and likewise define $c^{o p t}, p^{o p t}$ for each $\tilde{S}_{i j}$. Thus, we have two instances of the knapsack problem, namely $I_1=(\mathscr{C}, k, l, c, p, l)$ and $I_2=\left(\mathscr{C}^{o p t}, k, l, c^{\text {opt }}, p^{o p t}, l\right)$. 

For convenience, let's assume that $j$ represents the budget that should be spent for layer $G_i$, as determined by any MCKP solver. In that case, $S_{ij}$ and $\tilde{S}_{ij}$ can be used to represent the output, where $S_{ij}$ represents the approximate solution and $\tilde{S}_{ij}$ represents the unknown optimal solution. Since, our algorithm $\mathcal{A}$ is greedy-style, each element of $C_i$ has approximation ratio $(1-1/e)$. Thus, the value of the approximate solution $Otp_{S_{i j}}$ decided by any MCKP solver for each layer $G_i$ also has $(1-1/e)$ as follow:

\begin{equation}
\begin{aligned}
(1-1/e) O p t_{\tilde{S}_{ij}} & = (1-1/e) \sigma\left(\tilde{S}_{ij}\right) \\
& \leq \sigma\left(S_{ij}\right) \\
& \leq \sigma\left(S_{ij}\right) \\
& \leq O p t_{S_{ij}}
\end{aligned}
\end{equation}

 From there, we can also extend the analysis to get an approximation ratio for $\mathscr{G}$. Let, $S=\bigcup_{i=1}^k S_{ij}$ and $\tilde{S}=\bigcup_{i=1}^k \tilde{S}_{ij}$ represent the final solutions for MCKP instances $I_1$ and $I_2$, respectively. We have:

\begin{equation}
\begin{aligned}
(1-1/e) O p t_{\tilde{S}} & = (1-1/e) \sum_{i=1}^{k} \sigma\left(\tilde{S}_{ij}\right) \\
& \leq \sum_i \sigma\left(S_{ij}\right) \\
& \leq \sum_i \sigma\left(S_{ij}\right) \\
& \leq O p t_{S}
\end{aligned}
\end{equation}

Thus, proving the lemma.


\textbf{B.4. Approximation Ratio In The Worst Case}

\textbf{Theorem 1 }(Approximation Ratio In The Worst Case) Suppose the propagation $\sigma_i$ on each layer of the multiplex is submodular, the optimal policy $\pi^*$ will find a solution $\hat{S}$ for multiplex network $\mathscr{G}$ with an approximation ratio of $\frac{(1-\epsilon)(1-1/e)}{(o+1) k}$.

\textbf{Proof.} We start by assuming that the model $\sigma_i$ on each layer $G_i$ satisfies the Generalized Deterministic Submodular (GDS), as shown in Definition 6. If all $\sigma_i$ of all $G_i$ satisfy the Generalized Deterministic Submodular (GDS) property, then $\sigma$ is submodular (Kuhnle et al., 2018). 


Let $\hat{S}=\bigcup_{i=1}^k S_i$ be the solution returned by the MIM-Reasoner for the multiplex network $\mathscr{G}$, where $S_i$ represents the seed set found on each layer $G_i$ using policy $\pi^*$.
Let $\Tilde{S}$ be the unknown optimal solution for the multiplex network $\mathscr{G}$, and $\sigma(\hat{S})^i$ be the expected activation under $\sigma$ in layer $G_i$. Note that, $\sigma(\hat{S})^i$ only counts duplicated nodes as a single instance within layer $G_i$, rather than adding up all instances. Based on these definitions, we can state the following inequality:

\begin{equation}
\sigma\left(\Tilde{S}\right) \leq \sum_{i=1}^k \sigma\left(\Tilde{S}\right)^i    
\end{equation}

From this, we define $O$ as the set of native overlapping nodes. This set excludes isolated vertices that are added by other layers. Given the set $O$ of native overlapping nodes, it is established that:

\begin{equation}
    \sigma\left(\Tilde{S}\right)^i \leq \sigma_i\left(\Tilde{S} \cup O\right)
    \label{eqn:ovelapping1}
\end{equation}

Equation \ref{eqn:ovelapping1} arises because $\sigma(\Tilde{S})^i$ considers duplicated nodes as a single instance, while $\sigma_i(\Tilde{S} \cup O)$ counts all duplicated nodes. Moreover, any node in $G_i$ can be activated by the model $\sigma_i$ receiving seed nodes in $\Tilde{S} \cap G_i$ or native overlapping nodes $O$. In addition, because of submodularity of $\sigma_i$, we have:

\begin{equation}
    \sigma_i\left(\Tilde{S} \cup O\right)\leq \sigma_i\left(\Tilde{S}\right)+\sigma_i(O)
    \label{eqn:ovelapping2}
\end{equation}


Recall that in Lemma 2, $Opt_{S}$, $Opt_{\Tilde{S}}$ denotes the value of MCKP on instance $I_1$ and $I_2$, respectively. We also know that $\pi^*$ will select node greedily based on Lemma 1. Let $\sigma(\hat{S}^{\pi^*})$ denote the total spreading in multiplex network $\mathscr{G}$ of the solution $\hat{S}$ sampled by optimal policy $\pi^*$. In phase two, solution $S$ obtained from phase one is tuned by optimal policy $\pi^*$ to become $\hat{S}^{\pi^*}$, then:

\begin{equation}
    \sigma(\hat{S}^{\pi^*}) \geq \sigma(S) \geq (1-\epsilon) O p t_{S}
\end{equation}

Here, in the context of this work, $(1-\epsilon)$ with $\epsilon > 0$ represents the approximation ratio for the nearly exact solution found by the MCKP solver (Chandra et al., 1976) that we employ. In addition, for any set $S$ (could be optimal or just feasible) with size at most $l$ and any fixed layer $i$, we base on Lemma 2 to have :
\begin{equation}
\frac{1}{(1-\epsilon) (1-1/e)} \sigma(\hat{S}^{\pi^*}) \geq O p t_{\Tilde{S}} \geq \sigma_i(S)
\end{equation}
By Lemma 2, and since $\sigma_i(S)$ is the value of a feasible solution to the multiplex network instance in layer $G_i$, we have the above inequality.
Combining the inequalities, we get:
\begin{equation}
\begin{aligned}
\sigma\left(\Tilde{S}\right) & \leq \sum_{i=1}^k \sigma\left(\Tilde{S}\right)^i \\
& \leq \sum_{i=1}^k \sigma_i\left(\Tilde{S}\right)+\sum_{i=1}^k \sigma_i(O)\\
& \leq \frac{k}{(1-\epsilon) (1-1/e)} \sigma(\hat{S}^{\pi^*})+\sum_{i=1}^k \sigma_i(O) \\
& \leq \frac{k}{(1-\epsilon) (1-1/e)} \sigma(\hat{S}^{\pi^*})+\sum_{v \in O} \sum_{i=1}^k \sigma_i(v) \\
& \leq \frac{k}{(1-\epsilon) (1-1/e)} \sigma(\hat{S}^{\pi^*})+\frac{o k}{(1-\epsilon) (1-1/e)} \sigma(\hat{S}^{\pi^*}) \\
& \leq \frac{(o+1) k}{(1-\epsilon) (1-1/e)} \sigma(\hat{S}^{\pi^*}) .
\end{aligned}
\label{eqn:combinequa}
\end{equation}
We can rewrite this inequality as: 

\begin{equation}
    \sigma(\hat{S}^{\pi^*}) \geq \frac{(1-\epsilon) (1-1/e)}{(o+1) k}\sigma\left(\Tilde{S}\right)
\end{equation}

Thus, MIM-Reasoner has approximation ratio $ \frac{(1-\epsilon) (1-1/e)}{(o+1) k}$.

\textbf{B.5. Approximation Ratio In The Best Case}




\textbf{Theorem 2 }(Accuracy Gap In The Best Case). Assume the $\pi^*$ can avoid reactivating all the activated nodes, the spread of solution given by optimal policy $\pi^*$ is at least: $\sigma\left(\hat{S}^{\pi^*}\right) \geq \frac{(1-\epsilon)(1-1/e)}{k+o} \sigma(\tilde{S})$.

\textbf{Proof.} Recall that the optimal policy $\pi^*$ will avoid reactivating nodes that have already been activated, starting from layer $2$ up to layer $k$. We modify the proof of Theorem 1 (Equation $\ref{eqn:combinequa}$) to have:

\begin{equation}
\begin{aligned}
\sigma\left(\Tilde{S}\right) & \leq \sum_{i=1}^k \sigma\left(\Tilde{S}\right)^i \\
& \leq \sum_{i=1}^k \sigma_i\left(\Tilde{S}\right)-\sum_{i=2}^k \sigma_i(O)+\sum_{i=1}^k \sigma_i(O)\\
& \leq \frac{k}{(1-\epsilon) (1-1/e)} \sigma(\hat{S}^{\pi^*}) + \sigma_i(O) \\
& \leq \frac{k}{(1-\epsilon) (1-1/e)} \sigma(\hat{S}^{\pi^*}) + \sum_{v \in O}\sigma_i(v) \\
& \leq \frac{k}{(1-\epsilon)(1-1 / e)} \sigma\left(\hat{S}^{\pi^*}\right)+\frac{o}{(1-\epsilon)(1-1 / e)} \sigma\left(\hat{S}^{\pi^*}\right) \\
& \leq \frac{k+o}{(1-\epsilon)(1-1 / e)} \sigma\left(\hat{S}^{\pi^*}\right) 
\end{aligned}
\end{equation}

We can rewrite this inequality as: 

\begin{equation}
    \sigma(\hat{S}^{\pi^*}) \geq \frac{(1-\epsilon) (1-1/e)}{k+o}\sigma\left(\Tilde{S}\right)
\end{equation}

Therefore, the theorem is proved.

\textbf{B.6. Approximation Ratio In The General Case}

\textbf{Theorem 3 }(Accuracy Gap In General Case). Assume the $\pi^*$ can avoid reactivating the activated nodes by other layers partially. Thus, with $\beta \in[0,1]$, the spread of solution given by optimal policy $\pi^*$ is at least: $\sigma\left(\hat{S}^{\pi^*}\right) \geq \frac{(1-\epsilon)(1-1 / e)}{(k-1) \beta o+o+k} \sigma(\tilde{S})$.

\textbf{Proof.} Given $\beta \in [0,1]$ representing the percentage of nodes that cannot be successfully avoided reactivation by the policy $\pi^*$, we have:

\begin{equation}
\begin{aligned}
\sigma\left(\Tilde{S}\right) & \leq \sum_{i=1}^k \sigma\left(\Tilde{S}\right)^i \\
& \leq \sum_{i=1}^k \sigma_i\left(\Tilde{S}\right)-(1-\beta)\sum_{i=2}^k \sigma_i(O)+\sum_{i=1}^k \sigma_i(O)\\
& \leq \frac{k}{(1-\epsilon) (1-1/e)} \sigma(\hat{S}^{\pi^*}) + \sigma_i(O) -(1-\beta)\sum_{i=2}^k \sigma_i(O)+\sum_{i=2}^k \sigma_i(O) \\
& \leq \frac{k}{(1-\epsilon) (1-1/e)} \sigma(\hat{S}^{\pi^*}) + \sum_{v \in O} \sigma_i(v)  +\beta\sum_{v \in O} \sum_{i=2}^k \sigma_i(v)\\
& \leq \frac{k}{(1-\epsilon)(1-1 / e)} \sigma\left(\hat{S}^{\pi^*}\right) +\frac{o}{(1-\epsilon)(1-1 / e)} \sigma\left(\hat{S}^{\pi^*}\right)+\frac{o(k-1)\beta}{(1-\epsilon) (1-1/e)} \sigma(\hat{S}^{\pi^*})\\
& \leq \frac{k+o+o(k-1)\beta}{(1-\epsilon)(1-1 / e)} \sigma\left(\hat{S}^{\pi^*}\right) 
\end{aligned}
\end{equation}

Thus, proving the theorem.

\textbf{B.7. Time complexity of structure learning for PGMs}

\textbf{Lemma 3 } (PGMs's Time complexity). The time complexity of structure learning for $\mathcal{G}^{\text {prev }}$ after $k$ selection step is $|\mathcal{Y}|^2 \cdot(k-1)$. 

\textbf{Proof.} In the Structure Learning step, given a set of representative nodes $\mathcal{Y} = [v_1, \ldots, v_q]$, we calculate pairwise mutual information for all $\frac{q(q-1)}{2}$ pairs and employ Kruskal's algorithm to construct a Minimum Weight Spanning Tree (MWST). The algorithm constructs the tree one edge at a time, in decreasing order of weights. The running time of this step is $\mathcal{O}(|\mathcal{Y}|^2)$ for $|\mathcal{Y}| = q$ variables, as it needs to consider all $\frac{q(q-1)}{2}$ edges. In addition, at each layer selection (except for the final layer selection), we have to train a PGM. Therefore, the total number of PGMs after training with a multiplex network consisting of $k$ layers will be $k-1$. Each PGM has a complexity is $\mathcal{O}(|\mathcal{Y}|^2)$ and we have $k-1$ PGMs after training process. Thus, the time complexity of structure learning after $k$ selection step is $O(|\mathcal{Y}|^2 \cdot(k-1))$.








\textbf{B.8. Time complexity of MIM-Reasoner}

\textbf{Theorem 4 } (Time complexity of MIM-Reasoner). The time complexity of the Budget Allocation is: $\max _{h \in k} t c\left(A, G_h, l\right)+(k l)^{\lceil 1 / \epsilon-1\rceil} \log k$ and the time complexity of Relation RL Optimization is : $O\left(|\mathcal{Y}|^2 \cdot(k-\right.$ $1)+\mathcal{Q})$ where $\mathcal{Q}$ is number of step for policy $\pi$ converge to optimal.


\textbf{Proof.} Given the $V_i^g$ predicted by GAT model $\mathcal{I}(.)$ for each layer $G_i$, MIM-Reasoner runs algorithm A in parallel $k$ times with the search space $V_i^g \in G_i$ and then utilizes the $(1-\epsilon)$ Multiple Choice Knapsack Problem (MCKP) solver. Let's denote the layer that takes the longest time to run as $G_h$. If $O\left(\textit{tc}\left(A, G_h, l\right)\right)$ represents the time complexity of algorithm A on $l$ seed nodes with graph $G_h$, then the time complexity of the $(1-\epsilon)$ Multiple Choice Knapsack Problem is determined by $O\left((k l)^{\lceil 1 / \epsilon-1\rceil} \log k\right)$. Therefore, we have an overall time complexity for the Budget Allocation phase:

\begin{equation}
    O\left({\underset{h\in k}{\max }} \textit{ tc}\left(A, G_h, l\right)+(k l)^{\lceil 1 / \epsilon-1\rceil} \log k\right)
\end{equation}

In the second phase, our time complexity comes from the time complexity of PGMs, and the total number of training step for policy $\pi$ become optimal. Let denotes $\mathcal{Q}$ as a total number of training steps for policy $\pi$ becomes optimal. Based on Lemma 3, we have time complexity for PGMs for multiplex with $k$ layers is $|\mathcal{Y}|^2 \cdot(k-1)$. Thus, the time complexity for the Relation RL Optimization phase is:

\begin{equation}
O\left(|\mathcal{Y}|^2 \cdot(k-\right.1)+\mathcal{Q})
\end{equation}

\textbf{C EXPERIMENT DETAILS}

We conducted our experiments on a machine equipped with an Intel(R) Core i9-13900k processor, 128 GB RAM, and two Nvidia RTX 4090 GPUs with 24GB VRAM each.

\textbf{C.1 Experimental Analysis}

\textbf{Synthetic Multiplex Network.} We compare MIM-Reasoner with other baselines on a synthetic multiplex network with 5,000 nodes and 25,000 edges, considering different overlapping rates and the number of layers. Each layer is a random graph generated using the Erdos-Renyi algorithm, with the overlapping percentage (30 $\%$, 50 $\%$, 70 $\%$) calculated based on the layer with the highest number of nodes. In cases where the number of overlapping users exceeds the number of nodes in a layer, we create isolated nodes to maintain the correct number of overlaps. For the propagation models, we consider both the Independent Cascade (IC) model and the Linear Threshold (LT) model. In the IC model, the propagation probability for each edge in a layer is determined as 1 divided by the degree of the target node. Additionally, in the LT model, the propagation threshold for each node in a layer is randomly assigned in the range $[0.5, 0.9]$.

\textbf{Real World Multiplex Network.} We evaluate MIM-Reasoner and other methods on four real-world multiplex networks: 1) Celegans (Stark et al., 2006): This network consists of 6 layers with 3879 nodes and 8191 edges; 2) Drosophila (Stark et al., 2006; De Domenico et al., 2015): The Drosophila network comprises 7 layers, 8215 nodes, and 43,366 edges; 3) Pope-Election (Domenico and Altmann, 2020): The Pope-Election network includes 3 layers, with 2,064,866 nodes and 5,969,189 edges. The link of these dataset can be found in this link: https://manliodedomenico.com/data.php. Meanwhile, the Twitter-Foursquare network (Shen et al., 2012) has 2 layers, with 93269 nodes and 17,969,114 edges. It can be found in this link: "https://url1.io/s/Vd3YD".

\begin{table*}[ht]
\caption{\textit{The synthetic network used for evaluation consists of 5000 nodes and 25,000 edges. It is important to note that while the initial multiplex network has 5,000 nodes and 25,000 edges, the number of nodes and edges can vary depending on the overlapping user percentage or the number of layers employed. This is because, If a vertex does not exist in some other layer, we can simply add it as an isolated vertex.}}

\centering
$
\begin{array}{|l|r|r|r|r|r|r|r|}
\hline & \text { 3 Layers } & \text {4 Layers} & \text{5 Layers} & \text{6 Layers} & \text{7 Layers} & \text{8 Layers} & \text{9 Layers} \\
\hline \text { Layer 1 node count } & 500 & 500 & 200 & 200 & 100 & 100 & 100 \\
\hline \text { Layer 2 node count } & 2000 & 1000 & 600 & 400 & 200 & 200 & 200 \\
\hline \text { Layer 3 node count } & 2500 & 1500 & 1000 & 600 & 400 & 300 & 300 \\
\hline \text { Layer 4 node count } & 0 & 2000 & 1400 & 800 & 600 & 500 & 400 \\
\hline \text { Layer 5 node count } & 0 & 0 & 1800 & 1200 & 800 & 600 & 500 \\
\hline \text { Layer 6 node count } & 0 & 0 & 0 & 1800 & 1200 & 800 & 600 \\
\hline \text { Layer 7 node count } & 0 & 0 & 0 & 0 & 1700 & 1000 & 700 \\
\hline \text { Layer 8 node count } & 0 & 0 & 0 & 0 & 0 & 1500 & 800 \\
\hline \text { Layer 9 node count } & 0 & 0 & 0 & 0 & 0 & 0 & 1400 \\
\hline \text { Total Nodes } & 7500 & 8000 & 9000 & 10800 & 11900 & 12000 & 12600\\
\hline \text { Total Edges (30 $\%$ case)} & 26500 & 26800 & 27160 & 27700 & 28060 & 28150 & 28360\\
\hline \text { Total Edges (50 $\%$ case)} & 27500 & 28000 & 28600 & 29500 & 30100 & 30250 & 30600\\
\hline \text { Total Edges (70 $\%$ case)} & 28500 & 29200 & 30040 & 31300 & 32140 & 32350 & 32840\\
\hline
\end{array}
$
\end{table*}

\textbf{C.1.1 Training Time.} 

When running the algorithms on both synthetic and real-world datasets, we observed the same characteristics in terms of training time for all the methods. First, as the number of layers increases, the overall network becomes more complex with a higher number of connections (Table 1). Conversely, each layer becomes simpler. This explains the experimental results observed in methods that operate on the whole network, such as ISF, DeepIM, and ToupleGDD, which exhibit increasing running times (training time for ML-based methods) as the number of layers increases and take longer compared to parallel algorithm operating on individual layers like KSN or MIM-Reasoner model. 

In contrast to approaches that operate on the whole graph, network decomposition methods that operate on individual layers tend to exhibit decreased running time (for the CO algorithm) or training time (for ML-based methods). This reduction in time can be attributed to the fact that as the number of layers increases, each layer becomes simpler since the number of edges is divided among the layers, independent of the number of connections (edges) between overlapping users. Therefore, it is understandable that the running time of KSN or the training time of MIM-Reasoner tends to decrease as the number of layers increases, distinguishing them from other methods. Another interesting observation is related to the training of ML-based approaches. In the case of MIM-Reasoner, it requires a lightweight model with only 63,960 parameters in our setting. In contrast, ToupleGDD and DeepIM have significantly larger models with 925,090 and 9,238,780 parameters, respectively. Consequently, the training time for each optimization step in DeepIM and ToupleGDD is much slower compared to MIM-Reasoner due to the increased complexity and parameter count of their models.

\textbf{C.1.2 Inference Time.} 

For both synthetic and real datasets, DeepIM shows faster inference time compared to other methods because it predicts end-to-end solutions directly. On the other hand, ToupleGDD, with the giant model, has to infer the seed nodes step by step, resulting in longer inference times. MIM-Reasoner has a similar inference mechanism to ToupleGDD, but it leverages batch inference to parallelize the solution inference process for each layer of the multiplex network. This approach combines the states of each layer to create a batch, allowing for efficient computation. As a result, the majority of the time is spent on the layer that requires the highest budget when making inference using MIM-Reasoner.

\textbf{C.1.3 Propagation Performance.} 

ISF is a greedy-based algorithm with an approximation ratio of (1 - 1/e) under the GDS property, making it the most consistent algorithm among the five methods in terms of algorithm quality when tested on small-scale synthetic graphs. However, applying ISF to large real-world multiplex networks is computationally challenging due to the need for multiple propagation simulations. Meanwhile, KSN is a parallel algorithm that finds solutions for each layer independently, making it the fastest algorithm. However, it does not consider the activation of nodes in other layers through overlapping nodes, resulting in a smaller approximation ratio as the number of overlapping nodes increases. This limitation makes its solution worse compared to other methods in both synthetic and real multiplex networks.

DeepIM utilizes deep graph representation learning and optimization in continuous space to discover important seed sets in large complex graphs with high accuracy and efficiency. However, DeepIM may suffer from instability and convergence issues when dealing with more complex networks, leading to longer training times or poorer performance. In large multiplex networks such as Twitter-Foursquare or Pope Election, training DeepIM can take a very long time for the feature loss to converge. If the feature loss has not converged and only the reconstruction loss has converged, stopping the training process would result in a very poor solution. On the other hand, ToupleGDD is a well-designed solution that utilizes Deep Reinforcement Learning (DRL) for optimization problems. It has demonstrated effectiveness in experiments with both synthetic and real-world datasets. However, one limitation of ToupleGDD is its performance in large search spaces, where it can suffer from sample efficiency problems. This means that it may not consistently provide good results due to the challenges of exploring and finding optimal solutions in such an expansive large multiplex graph with limited training time steps.

Our proposed approach, MIM-Reasoner, decomposes the multiplex network into separate layers, effectively reducing the search space and observation space, even in large real-world multiplex networks such as Twitter-FourSquare or Pope-Election. This layer-wise approach allows the policy to focus on and explore the unique characteristics of each layer more efficiently, leading to improved learning and optimization within the multiplex network. Additionally, MIM-Reasoner employs Probabilistic Graphical Models (PGMs) to analyze how nodes are influenced by different layers within each layer. This utilization of PGMs enables MIM-Reasoner to find effective solutions for the MIM problem while mitigating the impact of overlapping nodes, addressing a limitation of methods like KSN. Consequently, MIM-Reasoner consistently provides good results in both synthetic and real multiplex networks. The scalability and generalizability of MIM-Reasoner make it applicable to various multiplex network scenarios. Its ability to handle large real-world multiplex networks while still achieving good performance demonstrates its strength in solving MIM problems.

\textbf{C.2 Hyperparameters and MIM-Reasoner Implementations}

\textbf{C.2.1 Hyperparameters} For policy training, we utilize Proximal Policy Optimization (PPO) as it offers stability and ease of implementation. The hyperparameters for the Graph Attention Network (GAT) and Policy model are depicted in Table 2. Note that the hyperparameters of PPO are based on the original paper.

\begin{table*}
\caption{\textit{Hyperparameters for Reinforcement Learning framework and Graph Attention Network}}
\centering
$
\begin{array}{lc}
\hline \text { Hyperparameter } & \text { Value } \\
\hline \text { Learning rate for Actor Model } & 0.0003 \\
\text { Learning rate for Critic Model } & 0.001 \\
\text { Learning rate for GAT Model } & 0.01 \\
\text { Optimizer } & \text { Adam (Kingma \& Ba, 2015) } \\
\text { Total epoch per update } & 8 \\
\text { Update time step  } & 600  \\
\text { Minibatch size } & 256 \\
\text { Discount factor } \gamma & 1 \\
\text { Layers for GAT Model } & \text { GATConv } \\
\text { Layers for Actor-Critic Model } & \text { Fully Connected Layer } \\
\text { Activation function for GAT Model } & \text { Elu } \\
\text { Activation function for Actor-Critic Model } & \text { Tanh } \\
\text { Target network smoothing coefficient } & 1024 \\
\text { Entropy coefficient } & 0.01 \\
\text { Lambda } & 0.95 \\
\text { PPO Epsilon } & 0.2 \\
\text { Gradient Norm } & 0.5 \\
\hline
\end{array}
$
\end{table*}

\vfill

\end{document}